\documentclass[manuscript,10pt,screen]{acmart}

\AtBeginDocument{%
  }

\copyrightyear{2025}
\acmYear{2025}
\setcopyright{acmlicensed}
\acmConference[ICS '25]{2025 International Conference on Supercomputing}{June 8--11, 2025}{Salt Lake City, UT, USA}
\acmBooktitle{2025 International Conference on Supercomputing (ICS '25), June 8--11, 2025, Salt Lake City, UT, USA}
\acmDOI{10.1145/3721145.3725766}
\acmISBN{979-8-4007-1537-2/2025/06}

\renewcommand\footnotetextcopyrightpermission[1]{} % Removes footnote with conference info
\acmSubmissionID{288}

\usepackage{xcolor}
\usepackage[T1]{fontenc}
\usepackage{tikz}
\usepackage{rotating}  
\usetikzlibrary{decorations.pathreplacing}  
\usepackage{subcaption} 
\usepackage{listings}
\usetikzlibrary{decorations}
\definecolor{listinggreen}{rgb}{0,0.6,0}
\definecolor{listingkeywordcolor}{rgb}{1.0,0.4,0.0}
\definecolor{listinglightgray}{rgb}{0.8863,0.8863,0.8863}

\newcommand{\yes}{{\textbf{\scriptsize \checkmark}}}
\newcommand{\no}{{\textbf{\scriptsize $\times$}}}

\newcommand{\speedupOverTiramisu}{$2.02\times\ $}
\newcommand{\speedupOverPluto}{$3.36\times\ $}
\newcommand{\toolName}{Pearl}
\begin{document}
%%
%% The "title" command has an optional parameter,
%% allowing the author to define a "short title" to be used in page headers.
\title{\toolName : Automatic Code Optimization Using Deep Reinforcement Learning}

\author{Djamel Rassem Lamouri}
\email{dl5133@nyu.edu}
\affiliation{%
  \institution{New York University Abu Dhabi}
  \city{Abu Dhabi}
  \country{UAE}
}

\author{Iheb Nassim Aouadj}
\email{ia2280@nyu.edu}
\affiliation{%
  \institution{New York University Abu Dhabi}
  \city{Abu Dhabi}
  \country{UAE}
}

\author{Smail KOURTA}
\email{sk10691@nyu.edu}
\affiliation{%
  \institution{New York University Abu Dhabi}
  \city{Abu Dhabi}
  \country{UAE}
}
\author{Riyadh Baghdadi}
\email{baghdadi@nyu.edu}
\affiliation{%
  \institution{New York University Abu Dhabi}
  \city{Abu Dhabi}
  \country{UAE}
}
\renewcommand{\shortauthors}{Djamel R. Lamouri, Nassim I. Aouadj, Smail Kourta and Riyadh Baghdadi}

\begin{abstract}
Compilers are crucial in optimizing programs and accelerating their execution, particularly for compute-intensive tasks such as training deep learning models and conducting physics simulations.
However, optimizing programs automatically using compilers is not trivial.
Recent work has attempted to use reinforcement learning (RL) to solve this problem.
It has limitations though. Current methods either do not support the optimization of general loop nests or can only be used to optimize loop nests seen during training.
In this paper, we propose \toolName{}, a novel framework that uses deep reinforcement learning to automate compiler code optimization. It uses an RL agent to select the sequence of code optimizations a compiler should apply to make the input code run faster. This agent can optimize general loop nests (i.e., it is not domain-specific) and can generalize to programs unseen during training.
To enable the optimization of general loop nests, we propose a novel representation of the action space that allows the RL agent to select on which part of the loop nest a given code optimization should be applied.
One of the main challenges that hinder the development of RL agents for optimizing general loop nests is the fact that this task is data-intensive, with each experiment taking weeks. To avoid this problem and enable fast training of the proposed RL agent, we propose two methods: 1) execution time and legality check memoization; and 2) actor-critic pre-training.
We implement our approach in Tiramisu, a state-of-the-art polyhedral compiler designed for accelerating compute-intensive programs. Our approach streamlines the optimization process and offers performance improvements compared to existing methods.
To the best of our knowledge, \toolName{} is the first RL-based system to support general programs composed of loop nests manipulating tensors while still being able to generalize to programs unseen during training. It is also the first to support the class of polyhedral optimizations, a class of advanced loop nest optimizations.
We evaluate \toolName{} on a set of benchmarks, and demonstrate competitive performance improvements over state-of-the-art compilers. Notably, \toolName{} achieves a geometric mean speedup of \speedupOverTiramisu compared to Tiramisu and \speedupOverPluto compared to Pluto.
\end{abstract}

%% Keywords. The author(s) should pick words that accurately describe
%% the work being presented. Separate the keywords with commas.
%% A "teaser" image appears between the author and affiliation
%% information and the body of the document, and typically spans the
%% page.

%\received{20 February 2007}
%\received[revised]{12 March 2009}
%\received[accepted]{5 June 2009}

%%
%% This command processes the author and affiliation and title
%% information and builds the first part of the formatted document.
\maketitle

\section{Introduction}
Writing fast and efficient code is a challenging task that requires significant expertise. This is especially true for compute-intensive fields such as deep learning and scientific computing. Optimizing compute-intensive programs can significantly reduce execution time, often achieving speedups by orders of magnitude. However, manual code optimization is time-consuming, error-prone, and demands expertise, making automated compiler optimizations increasingly crucial.

The most significant part of the execution time of compute-intensive programs is usually spent in loops. That is why loop optimization has received considerable attention in the context of code optimization. State-of-the-art compilers such as Tiramisu~\citep{baghdadi2019tiramisu},  Halide~\citep{ragan2013halide}, TVM~\citep{zheng2020ansor}, and Pluto~\citep{bondhugula2008pluto} all focus on applying loop transformations to accelerate the execution of programs. Nevertheless, the search space of these transformations is considerably large, hindering the usage of exhaustive search methods due to their impracticality.

Several techniques have been proposed to solve this challenge, including the leverage of integer linear programming (ILP) to find the best code optimizations~\cite{bondhugula2008pluto,bondhugula2008practical,sven2013}. Recently, state-of-the-art optimizing compilers have explored tree-search methods guided by a deep learning cost model, serving as a fitness function to evaluate code optimization candidates during search
%, to explore the search space of code optimizations
~\citep{adams2019halide,zheng2020ansor,baghdadi2021deep}. To shed light on the size of the search space, let's take the Tiramisu Autoscheduler~\citep{baghdadi2021deep} as an example. Its search space has an estimated number of $10^{170}$ candidates of loop optimizations~\citep{baghdadi2021deep}, where each candidate is a sequence of code optimizations along with their parameters. This is why the Tiramisu Autoscheduler imposes restrictions on how the search space is explored. For example, code optimizations are explored in a fixed order, and many optimizations are explored only once, which is known to be sub-optimal. The Tiramisu Autoscheduler uses beam search to explore the space, limiting its ability to perform global code optimization. In general, tree-search methods tend to restrict the search space to enable efficient exploration. For example, they might explore only a small number of code optimizations, explore code optimizations in a fixed order, and limit the number of times certain code optimizations are explored (e.g., explore them once only).
%These restrictions limit the ability of the Autoscheduler to find significant optimization opportunities.

To avoid the limitations of tree-search methods, recent state-of-the-art compilers have attempted to use reinforcement learning. HalideRL~\citep{pecenin2019optimization}, for example, uses PPO \cite{schulman2017proximal} to select code optimizations for image processing applications.
SuperSonic~\citep{huanting2022ss}, a framework for automating RL architecture design for code optimization, was also demonstrated by building an RL agent for the Halide compiler. However, neither HalideRL nor SuperSonic's Halide agents are built to generalize over programs unseen during training. In both cases, the RL agent is trained to optimize a set of programs, and then it is deployed to optimize the same programs. This design choice allows fast training of the RL agent since it is trained on a single program only, but it prevents its generality.
%deployed to optimize programs that are congruent with the training set.
Additionally, HalideRL has the limitation of being semi-automatic. It requires the user (developer) to specify the list of code optimizations and optimization parameters that will likely help optimize the code. The RL agent then discards the optimizations that are not useful from this list and adjusts the optimization parameters.

Other state-of-the-art systems focus on proposing an environment for automatic code optimization using reinforcement learning but do not propose an RL agent that automatically optimizes code. PolyGym~\citep{brauckmann2021reinforcement}, for example, introduced an RL environment to search for polyhedral code optimizations (affine code optimizations). It does not propose an RL agent though. It rather uses a random policy to demonstrate the effectiveness of its environment and action space. Such a random policy has the limitation of being slow as it has to randomly sample thousands of actions from the search space, and for each sampled action, it has to compile and run the optimized program to find the most effective ones. Proposing an RL agent that efficiently search for code optimizations in PolyGym was left for future work.

AutoPhase~\citep{huang2020autophase} is another example of using RL for code optimization. It proposes a framework that uses deep reinforcement learning to optimize programs for hardware synthesis (High-Level Synthesis). It only considers the problem of phase reordering (i.e., choosing the best order for the compiler passes). Phase ordering is a sub-problem of the larger problem of automatic code optimization. In automatic code optimization, the goal is to identify which optimizations to apply, in which order, on which part of the code each of them should be applied, and with which parameters. Chameleon~\cite{ahn2020chameleon} is another example of a compiler that uses reinforcement learning to find the best compiler transformations to accelerate the execution time of neural networks during deployment. However, it is not general. It is limited to the acceleration of deep learning models and does not cover the optimization of general loop nests, which limits its applicability. Optimizing programs with loops is a much harder problem since general loops can have diverse and complex structures and code patterns, unlike deep learning operators which comprise a limited set of operators with regular shapes and code patterns.

In this paper, we present \toolName{}, a deep reinforcement learning-based system for polyhedral code optimization. It supports the optimization of general loop nests and generalizes to programs unseen during training. \toolName{} avoids the limitations of tree-search methods and uses a deep policy network to predict the sequence of code optimizations to apply. We also propose a novel representation of the action space that enables the RL agent to select on which part of the loop nest a given code optimization should be applied.
Our technique explores a large action space that includes six loop transformations with their parameters. Unlike existing work, it supports a set of polyhedral code optimizations, it can generalize to unseen programs, and can be applied to any program that can be expressed as a sequence of loop nests and operates on tensors. Examples of types of computations that our RL agent supports include image processing, deep learning, linear algebra, stencil computations, and tensor operators.

During the development of the RL agent, one needs to train the RL agent repeatedly, to test different features and hypotheses and to fine-tune hyper-prameters. With each training taking weeks, and with the need for tens of experiments, the development of the RL becomes challenging. We also propose a set of techniques to mitigate this challenge. The main idea of our proposed technique is to store any value computed when training the RL agent, if it can be reused in a future training.
For example, when the RL agent picks a particular action, it has to check whether it is legal, and if so, it applies it by compiling the optimized program and then gets the reward by running the optimized program. Both of these operations are computationally expensive and constitute a significant part of the RL training time. We store the results of these operations (legality and reward of an action) in a dataset and reuse them in future trainings. In subsequent trainings, when the RL picks an action, we first check whether its legality or reward have already been computed in previous trainings. If so, we retrieve them directly, otherwise we compute them and add them to the dataset. We call this method \emph{execution time and legality check memoization} and it helps significantly in accelerating the training. We also propose another technique where we pre-train the actor-critic network of the RL agent with data collected in previous trainings. We call this technique \emph{actor-critic pre-training} and it helps also in improving the learning in the actor-critic neural networks.

Unlike HalideRL and SuperSonic's Halide agent, \toolName{} generalizes to programs unseen during training. The agent is trained on randomly generated programs and evaluated on a completely different set of benchmarks. It is also fully automatic. It does not require input from the user. In contrast with PolyGym, which only proposes an environment, we propose both an environment and a deep RL agent for that environment. In contrast to AutoPhase, which selects the best order for compiler passes, our goal is to tackle the more general problem of automatic code optimization, which includes selecting optimizations and their parameters, the order of applying them, and on which part of the code. Unlike Chameleon, our system is not limited to accelerating deep neural networks but is general to any program that can be expressed as a sequence of loop nests and operates on tensors.

We implement the proposed approach in the Tiramisu compiler, a state-of-the-art compiler~\cite{baghdadi2019tiramisu}, and evaluate it on a set of benchmarks from the fields of linear algebra, image processing, and scientific computing. We show competitive performance improvements compared to state-of-the-art compilers. Notably, our proposed approach achieves an overall geometric mean speedup of \speedupOverTiramisu  compared to Tiramisu and \speedupOverPluto compared to Pluto. You can find the full implementation of our method.\footnote{Code available at \url{https://github.com/Modern-Compilers-Lab/GNN_RL_Pretrain}}

In this paper, we make the following contributions:
\begin{itemize}
    \item We introduce \toolName{}, a deep reinforcement-learning system for polyhedral loop nest optimization.
    \item We propose a novel representation of the action space that allows the RL agent to optimize loop nests. It allows the RL agent to choose the most appropriate code optimizations for each part of the loop nest.
    \item To the best of our knowledge, \toolName{} is the first RL system that supports the optimization of programs composed of general loop nests and can still generalize to programs unseen during training.
    \item To the best of our knowledge, \toolName{} is also the first to propose an RL agent that supports the class of polyhedral optimizations.
    \item We implement and evaluate \toolName{} and show that it outperforms state-of-the-art.
    \item We release our dataset and make our code publicly available to the community.
\end{itemize}

\section{Related Work}

In this section, we provide an overview of state-of-the-art methods used by compilers for automatic code optimization (auto-scheduling). First, we present compilers that use a search-based method and a learned performance model for automatic code optimization. We then present early attempts to use reinforcement learning for automatic code optimization. Table~\ref{tab:related} shows a summarized comparison between these methods. Finally, we present other methods that do not rely on machine learning but rather use Integer Linear Programming (ILP) for automatic code optimization.

\begin{table}[tb]
    \centering
    %\footnotesize
    \
    \vspace{0.5cm}
    \setlength\tabcolsep{2pt}
    \caption{\textsc{Comparison with RL-based Systems.}}
    \begin{tabular}{|p{5cm}|c|c|c|c|c|c|c| }
        \hline
        
        \textbf{Features} & \textbf{\rotatebox{90}{\ \toolName{}\ }} & \textbf{\rotatebox{90}{\ PolyGym\ }} & \textbf{\rotatebox{90}{\ CompilerGym\ }} & \textbf{\rotatebox{90}{\ HalideRL\ }} & \textbf{\rotatebox{90}{\ AutoPhase\ }} & \textbf{\rotatebox{90}{\ Chameleon\ }} & \textbf{\rotatebox{90}{\ X-RLflow\ }} \\\hline

        \textbf{RL Environment}                     & \yes  & \yes  & \yes  & \yes & \yes & \yes & \yes \\\hline

        \textbf{RL Agent}                           & \yes & \no   & \no   & \yes & \yes & \yes & \yes \\\hline

        \textbf{Fully Automatic}                    & \yes  & \yes & \yes & \no & \yes & \yes & \yes \\\hline
        
        \textbf{Support Affine Transformations}   & \yes  & \yes & \no   & \no & \no & \no & \no \\\hline

        \textbf{Supports General Loop Nests}        & \yes  & \yes & \yes & \yes & \yes & \no & \no \\\hline

        \textbf{Optimize Unseen Programs}           & \yes  &   -  &   -  & \no  & \yes & \yes & \yes \\\hline

        \textbf{Auto-scheduling Framework}         & \yes  & \yes & \yes & \yes & \no & \yes & \yes \\\hline

    \end{tabular}
    \label{tab:related}
\end{table}

\paragraph{Search-based methods with cost models.}

This method was widely used due to its efficiency in exploring the large space of possible candidates. A heuristic search method is guided by a cost model to locate the best sequences of code optimizations. Tree-search methods were successfully used in Tiramisu~\cite{baghdadi2021deep,merouani2024looper,hakimi2023hybrid,liu2025data,mezdour2023deep}, Halide~\cite{adams2019halide}, and ProTuner~\cite{haj2020protuner}.
%which is built on Halide but uses MCTS which proved to be better than beam search.
Other work uses genetic and evolutionary algorithms to explore the search space~\cite{chen2018tvm, chen2019learning, zheng2023ansor} but they follow the same approach. RL-based methods surpass search-based methods in two ways. First, search-based methods are slower since they explore a large space of code optimizations and evaluate the best candidates within that space. As an example, Halide~\cite{adams2019halide} evaluates millions of candidates in the search space, while Tiramisu~\cite{baghdadi2021deep} evaluates thousands of candidates. In our proposed system, \toolName{}, the policy network selects the most appropriate sequence of code optimizations directly, without the need to explore a large space.
Second, many search-based methods constrain the order of exploring code optimizations to keep the size of the search space smaller, which leads to sub-optimal results.

\paragraph{RL-based methods.}

Recent attempts explored the use of reinforcement learning to solve the problem of choosing the right sequence of code transformations. In PolyGym~\cite{brauckmann2021reinforcement} and CompilerGym~\cite{cummins2022compilergym}, the authors propose only RL environments without implementing RL agents to optimize code, their main contribution is to show that their action space has potentially good optimizations to explore. They leave the implementation of an RL agent as future work.

Other works such as HalideRL~\cite{pecenin2019optimization}, AutoPhase~\cite{huang2020autophase} and SuperSonic~\cite{huanting2022ss} propose RL agents to optimize code.
HalideRL is not fully automatic. The user has to provide an initial set of code transformations. The HalideRL agent then discards transformations that are not useful and keeps only those that are useful. It then selects the best parameters for the useful transformations. In addition, HalideRL does not generalize to programs unseen during training. It is trained on a given program with multiple random data input sizes. Then during deployment, it is used to optimize that same program. This is different from our approach. Our RL agent is designed to generalize to programs unseen during training. We first train our RL agent on a large set of random programs. Once it learns how to optimize them, we then deploy it on new unseen programs and use it to optimize them.

SuperSonic~\cite{huanting2022ss} is a meta-optimizer that targets the problem of choosing the best RL algorithm and the best representation of states and actions, while AutoPhase~\cite{huang2020autophase} targets the problem of phase ordering, i.e., selecting the best order for compiler passes. It does not target the problem of auto-scheduling which we address in this paper. Phase ordering is only a sub-problem of the larger problem of auto-scheduling. In addition, AutoPhase targets HLS (High-level Synthesis) and does not target CPUs which we focus on.

All of the previous approaches, with the exception of PolyGym, are not designed for the class of affine transformations that we target (a.k.a., polyhedral transformations). Affine transformations allow advanced code optimizations for improving data locality and parallelism extraction and are hard to model due to their complexity and diversity.
A summarized comparison between our proposed RL and RL-based compiler frameworks is presented in Table~\ref{tab:related}.

\paragraph{Graph level optimization for deep learning.}

Compilers in this category target the application of code transformations in the domain of deep learning. With a higher level of abstraction, these compilers consider graph-level transformations on the deep learning computation graphs. They focus on transformations such as data layout transformations~\cite{liu2019optimizing}, operator fusion~\cite{chen2018tvm, zheng2021fusionstitching}, and auto-batching~\cite{looks2017deep}. Chameleon~\cite{ahn2020chameleon} uses reinforcement learning to find the best compiler transformations that accelerate the execution time of neural networks during deployment. REGAL~\cite{paliwal2020reinforced} targets only model parallelism on computational graphs to minimize execution time and memory peak usage. X-RLflow ~\cite{he2023xrlflowgraphreinforcementlearning} addresses the tensor graph superoptimisation problem using graph neural networks and reinforcement learning to perform neural network dataflow graph rewriting, which substitutes a subgraph one at a time. Our method is different from these in the sense that it does not operate on the high-level abstraction of computation graphs, but rather operates on a lower level. Because of that, our proposed RL agent is more general since it is not limited to deep learning computation graphs, but rather supports any program that can be expressed as a sequence of loop nests and operates on tensors. Types of computations that our RL agent supports include image processing, deep learning, linear algebra, stencil computations, tensor operators, etc.

\paragraph{Polyhedral Optimization with ILP} 

The polyhedral model~\cite{Feautrier2011} is a mathematical model for representing code and code transformations and is used in state-of-the-art compilers to apply complex code transformations and reason about their correctness~\cite{Iri88,feautrier_array_1988,wolf1991loop,lefebvre_automatic_1998,Qui00,thies_unified_2001,Darte_contraction_2005,bondhugula2008pluto,bondhugula_practical_2008, baghdadi2015PhD,baghdadi2015pencil,baghdadi2013pencil,baghdadi2019tiramisu,baghdadi2018tiramisu1,trifunovic_graphite_2010,polly,tobias_hexagonal_cgo13,Vasilache2018TensorCF,baghdadi2011speculation,merouani2020deep, pouchet.11.popl,vasilache2018tensor,baghdadi2020tiramisuDNNDenseSparse}.
As a method to solve the code optimization problem, Integer Linear Programming (ILP) was used by \cite{bondhugula2008pluto, bondhugula2008practical,sven2013} to explore the search space and find optimal solutions. The limitation of these ILP-based approaches is the lack of a precise cost model for predicting the performance of code optimizations. This limitation comes from the use of ILP which limits the cost function to a simple linear cost function. Because of this, more recent polyhedral compilers, such as Tiramisu, have switched to the use of search-based approaches along with a deep learning cost model for performance prediction.

\section{Background}

\subsection{Reinforcement Learning}
Reinforcement learning is a machine learning paradigm where an agent learns from interacting with an environment to maximize a cumulative reward \cite{sutton2018reinforcement}. In this work, we model the problem of automatic code optimization as a Markov Decision Process represented as a tuple $M = \langle S, A, P, R, \gamma \rangle$. In this model, $S$ is the set of all states while $A$ is the set of all actions. $P$ is the transition probability to a state $s^{'}$ given a the state $s$ and action $a$ where $P(s^{'}| s, a) = \mathbb{P}[S_t = s^{'} | S_{t-1} = s, A_{t-1} = a]$, and $R$ is the reward function that indicates the expected reward for a given state-action pair $R(s, a) = \mathbb{E}[R_t | S_{t-1} = s, A_{t-1} = a]$. The discount rate $\gamma \in (0, 1)$ determines the weight of future rewards in the agent's decision-making process \cite{sutton2018reinforcement}.  As the agent interacts with the environment, it learns a policy $\pi(a|s)$ that maximizes the cumulative reward. We are interested in policy-based RL algorithms. We train our agent using Proximal Policy Optimization \cite{schulman2017proximal}.   %Deep Reinforcement Learning (Deep RL) is a technique that involves using deep neural networks (DNN) with the RL, often to represent the policy.

\subsection{Graph Neural Networks}
In our work, we consider undirected attributed graphs $G(V, E, X)$ where $V$ is the set of nodes, $E$ is the set of edges, and $X \in \mathbb{R}^{|V|\times d}$ is an input matrix representing node features. $x_v \in \mathbb{R}^d$ denotes the features of node $v \in V$. Modern GNNs use the message passing paradigm \cite{gilmer2017neural} to update the node features ${h^{(k)}_v}$ in an iterative process where ${h^{(0)}_v} = x_v$. Message Passing Neural Networks (MPNNs) update ${h^{(k)}_v}$ as follows
\[
{h^{(k+1)}_v} = \phi^{k}({h^{(k)}_v}, \psi^{k}({h^{(k)}_v}, \{{h^{(k)}_u | u \in \mathcal{N}_v}\})),
\]
where $\mathcal{N}_v$ is the set of neighbors of node $v$, $\psi^k$ is a permutation invariant function that aggregates the representation of the neighborhood of the node $v$ into fixed size vector representation, and $\phi^k$ is the update function that takes the $k^{th}$ representation of node $v$ with the $k^{th}$ representation of the neighborhood to compute ${h^{(k+1)}_v}$. For graph-level prediction, we compute $h_{G}^{(k)}$ using another permutation invariant function that aggregates all node representations at iteration $k$ into a single vector.
\[
h_{G}^{(k)} = readout(\{h^{(k)}_v | v \in V\})
\]
We use Graph Attention Networks (GATv2) \cite{brody2022attentive}, a famous MPNN model based on the attention mechanism.

\subsection{Tiramisu}
We implement our method in the Tiramisu compiler \cite{baghdadi2019tiramisu}.
%Tiramisu offers a compiler with a language that allows the developer to write an algorithm and select which code optimizations to apply on code. This enables targeting multiple high-performance architectures using the same code base.
Tiramisu allows the user to express algorithms composed of sequences of loop nests and statements that manipulate scalars and tensors. Such algorithms dominate compute-intensive domains, including image processing, linear algebra, stencil computations, and deep learning.

Tiramisu provides a compiler and a DSL (Domain-Specific Language) embedded in C++ to represent code and its transformations. It provides two APIs: (a) an API for developers to write high-level architecture-independent algorithms, and (b) a second set of functions to describe how to optimize the algorithm. This separation enables the compiler to try a variety of code optimizations on the same algorithm.

In general, every Tiramisu program\footnote{Also referred to as Tiramisu function.} can be divided into two parts. The first part describes the algorithm as a function with inputs and outputs. Each program comprises a sequence of computations (a computation is a loop nest with a statement in its body). The second part of a Tiramisu program is reserved for specifying how the algorithm is optimized using a specific API.

\subsection{Polyhedral Access Relations (Access Matrices)\label{accessmatrices}}

In this section, we introduce the concept of polyhedral access relations, which are used in the polyhedral model to represent array accesses. These access relations are represented using matrices called access matrices (in other words, access matrices are a matrix representation of the access relations). We pass the access matrices as input to the RL deep learning models to represent array accesses. In the rest of this section, we will first explain the concept of access relations and then show how an access relation is represented using an access matrix.

Access relations are a set of read, write and may-write access relations that capture memory locations on which statements operate. They map statement execution instances to the array elements that are read or written by those instances.

\vspace{0.25cm}
\begin{lstlisting}[stepnumber=0,numbers=left,numberstyle={\tiny\tt},numbersep=5pt,escapechar=@,language=c]
for (i=1; i<=2; ++i)
  for (j=1; j<=2; ++j)
S:    A[i,j] = B[i,j];
\end{lstlisting}
\vspace{0.25cm}

In the previous example, the set of read-access relations is

$$ R_s = \{ S[i,j] \rightarrow B[i,j]\} $$

\noindent which means that the statement $S$ in iteration $i,j$ reads
the array element $B[i,j]$.

The set of write access relations is

$$ W_s = \{ S[i,j] \rightarrow A[i,j]$$

The read access relation $R_s$ can also be represented using a matrix as follows:

 $$
   R_s \left(
 	        \begin{array}{c}
 	        i\\
 	        j\\
 	        \end{array}
        \right) =
     \left(
 	\begin{array}{c c}
 	 1 & 0\\
 	 0 & 1\\
 	\end{array}
     \right)
      \left(
 	        \begin{array}{c}
 	        i\\
 	        j\\
 	        \end{array}
        \right)
 $$

\noindent where the matrix 
$$
        \left(
        \begin{array}{c c}
 	 1 & 0\\
 	 0 & 1\\
 	\end{array}
        \right)
$$
is the matrix representation of the access relation in this case. We call it, the access matrix. This access matrix is the representation that we pass as input to the RL deep learning models.

\begin{figure*}[h!t]
    \centering
    \includegraphics[width=12cm]{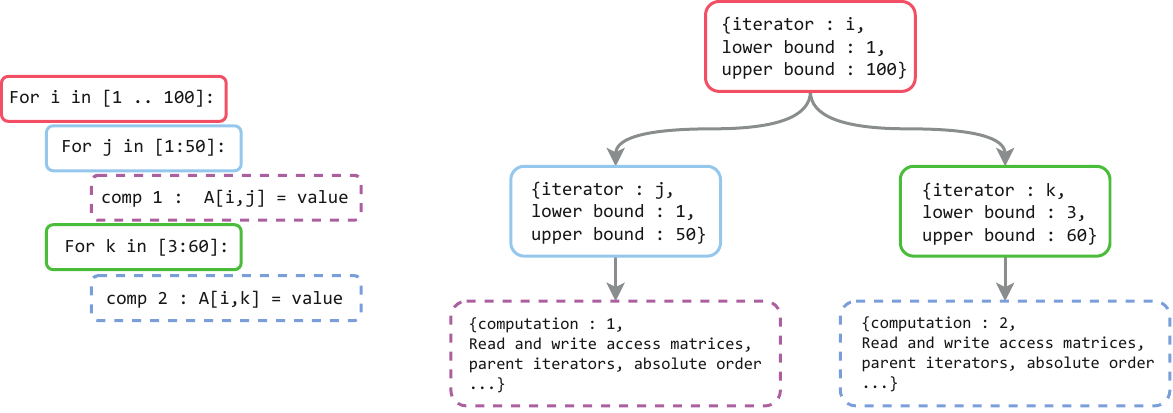}
    \caption{The construction of the Abstract Syntax Tree from a loop nest in a Tiramisu function. The nodes in the tree with dashed line rectangles represent statements or computations. The nodes with solid line rectangles represent iterators.}
    \label{fig:ast}
\end{figure*}

\begin{center}

\begin{figure}[H]
\centering
\tikzset{every node/.style={font=\scriptsize}}  % Makes all node text consistent and safe
\begin{tikzpicture}[
  roundnode/.style={circle, draw=gray, fill=white, thick, minimum size=3mm},
  rectnode/.style={rectangle, draw=gray, fill=white, thick, minimum size=3mm},
  vectornode/.style={rectangle, minimum size=5mm, inner sep=2mm},
]  

\node[roundnode, style={draw=white}] (gat1) at (-0.9,0.2) { \; GAT};
\node[roundnode, style={draw=white}] (gat2) at (0.4,-2.2) {GAT};

\node[roundnode] (x) at (-1.5,0) {\textbf{$X$}};
\node[roundnode] (h1) at (0,0) {\textbf{$H^{(1)}$}};
\node[roundnode] (h2) at (0,-3) {\textbf{$H^{(2)}$}};

\node[roundnode] (s1) at (1.5,0) {$\oplus$};
\node[roundnode] (s2) at (1.5,-3) {$\oplus$};

\node[roundnode] (hg1) at (3,0) {\textbf{$h_{G}^{(1)}$}};
\node[roundnode] (hg2) at (3,-3) {\textbf{$h_{G}^{(2)}$}};

\node[roundnode] (c) at (4.5,-1.5) {$\parallel$};
\node[roundnode] (g) at (6,-1.5) {$h_{G}^{f}$};

\draw[thick, color=black,->] (x) -- (h1);
\draw[thick, color=black,->] (h1) -- (h2);
\draw[thick, color=black,->] (h1) -- (s1);
\draw[thick, color=black,->] (h2) -- (s2); 
\draw[thick, color=black,->] (s1) -- (hg1);
\draw[thick, color=black,->] (s2) -- (hg2);
\draw[thick, color=black,->] (hg1) -- (c);
\draw[thick, color=black,->] (hg2) -- (c); 
\draw[thick, color=black,->] (c) -- (g); 

\end{tikzpicture}
\caption{The process of building $h_G^{f}$ where $\oplus$ produces two vectors, the first vector by summing node features $H^{(k)}$ and the second vector by applying element-wise maximization. These two vectors are concatenated to form $h_{G}^{(k)}$, representing aggregated graph features after $k$ message passing steps. The $\parallel$ symbol represents the final concatenation of vectors $h_{G}^{(k)}$ to construct $h_G^{f}$.}
\label{fig:gnn_layers_states}
\vspace{-0.5cm}
\end{figure}
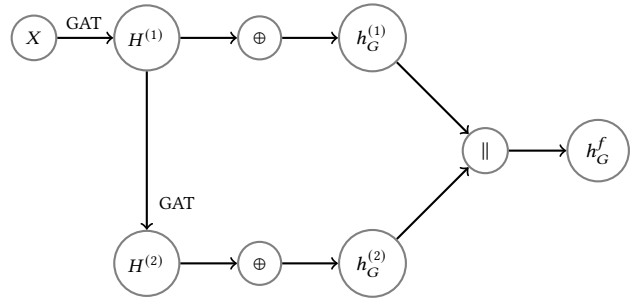

\end{center}
\section{Method Description} 

\subsection{States Representation} \label{subsec:states_repr}
Our work targets loop transformations since loops take most of a program's execution time. This aligns with the common practice in state-of-the-art compilers with automatic code optimization capabilities~\cite{baghdadi2019tiramisu,bondhugula2008pluto,ahn2020chameleon,ragan2013halide,adams2019halide,mallupudi2016halide,vasilache2018tensor,sven2013}. To represent loops written in Tiramisu code, we use an intermediate representation under the form of a tree known as an Abstract Syntax Tree (AST). A node $i$ in the tree can be either an iterator\footnote{We use the words iterator and loop interchangeably.} or a computation, and an edge $e_{ij}$ between two nodes $i$ and $j$ signifies that node $i$ is the parent of node $j$. Children nodes can be iterators or computations\footnote{We use the words computation and statement interchangeably.}, while parent nodes can only be iterators. Figure \ref{fig:ast} illustrates an example of how the AST is built from a given loop nest\footnote{Loop nest: set of loops where one loop is contained within another.}.

We build the node representation matrix $X$, and the graph edges $E$ from the AST. A row in the matrix $X$ represents the features of a single node in the AST. For the graph edges ($E$), we use simple edges with no attributes and keep the connections between the nodes as they originally existed in the AST. We map information from the AST nodes into fixed-size vectors for node features. The entire representation of nodes is detailed in Section~\ref{DetailedRepresentation}.

As illustrated in Figure \ref{fig:gnn_layers_states}, we pass $X$ through 2-layers of message passing using a GATv2 model and aggregate node representations at each layer using average and max pooling. Finally, we concatenate the aggregated representations from each layer of message passing to produce a final feature vector $h_{G}^f$ that represents the graph. We refer the reader to the background section for a more detailed description of how graph neural networks process their inputs.

\begin{figure*}[ht]
    \centering
    \includegraphics[width=13cm]{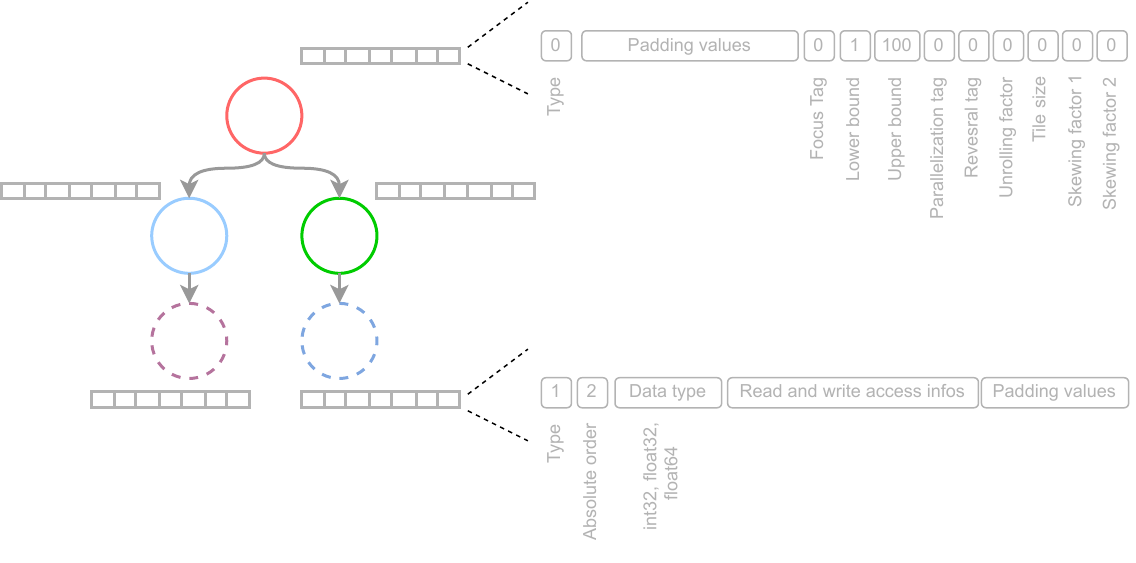}
    \caption{The construction of nodes features.}
    \label{fig:node_feats}
\end{figure*}
\subsubsection{Node Feature Representation} \label{DetailedRepresentation}

There are two types of nodes in the AST: iterators and computations. To make the difference between the two types, we use a vector of the same size to represent both types with different tags and padding. Using the same example as above, we will depict the details of each node representation in \autoref{fig:node_feats}.

As shown in \autoref{fig:node_feats}, the first column differentiates the representation of the iterator nodes from the computation nodes, followed by specific features for each type of node. For the iterators, the "Focus Tag" tells the agent that we are targeting this iterator for optimization, so the following action will potentially include that node. The other tags, like Parallelization and Reversal, represent whether or not those actions have transformed the iterator. In addition to the read and write access matrices (following the concept described in section \ref{accessmatrices}).  We stack those vectors to form the input node representation matrix $X$.

\subsection{Actions and AST branches} \label{sec:branch_action}

In this work, we consider the following loop transformations: 1) loop parallelization (to exploit multicore parallelism); 2) loop unrolling (which unrolls loop iterations to exploit instruction level parallelism); 3) loop tiling (to improve data locality); 4) loop skewing (which allows the extraction of outer parallelism and improves data locality);  5) loop interchange (which changes the order of loops within a loop nest to enable better data locality and parallelism); 6) loop reversal (which reverses the order of the loop iterations to enable better data locality and parallelism). Each of these transformations is applied to particular loops within a loop nest and many transformations have parameters. For example, loop tiling when applied on three loops, has the tuple $(T0, T1, T2)$ as a parameter where $T0, T1, T2$ are the tile sizes for each one of the three loops being tiled. They can take a value equal to the power of 2 and is between $2$ and $256$. The size of the search space covered by these transformations and their parameters is in the range of $10^{170}$~\cite{adams2019halide,baghdadi2021deep}.

At each time step $t$, the RL agent must determine the iterators impacted by the action $a_t$, where an action is defined by the tuple $(\mathcal{T},\mathcal{I},\mathcal{C},\mathcal{F})$, where $\mathcal{T}=$ \{Parallelization, Unrolling, Tiling, Skewing, Interchange, Reversal, Next\} is the set of loop transformations types in addition to "Next" which does not transform the loops but used to switch between branches, this action will be described in the next paragraph. The sets $\mathcal{I}$ and $\mathcal{C}$ represent the iterators and computations affected by $a_t$, respectively. Additionally, $\mathcal{F}$ is the set of transformation-specific parameters, such as tile sizes for Tiling. 

\begin{figure}[ht]
    \centering
    \includegraphics[width=6cm]{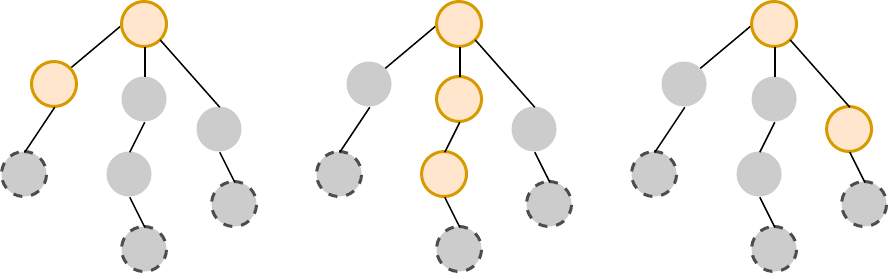}
    \caption{The graph at the left has 2 nodes colored in orange that represent the initial targeted branch. After applying loop transformations on the selected iterators, the agent uses the "Next" action to switch to the middle branch, as illustrated by the middle graph. The graph on the right represents the last branch. Choosing "Next" from that state will put an end to the episode and the optimization process.}
    \label{fig:branches}
\end{figure}

A given loop transformation is usually applied on a particular iterator within the loop nest. One might create an action space where each tuple (loop-transformation, AST-branch\footnote{A branch is a path from the root to a leaf in the AST}, iterator) becomes an action. This is not possible though, because the action space required in this case would be vast. This is mainly because ASTs of programs can take different forms with many branches and depths. To solve this challenge, the agent traverses the AST progressively, branch by branch. It starts at the leftmost branch of the tree as shown in  \autoref{fig:branches}, the agent chooses actions that target the iterators inside that branch and after selecting the "Next" action, the agent will target the next branch going from left to right to traverse loops by their order of appearing in the program. The episode ends when the agent is targeting the rightmost branch and the action "Next" is chosen.

% Figure:
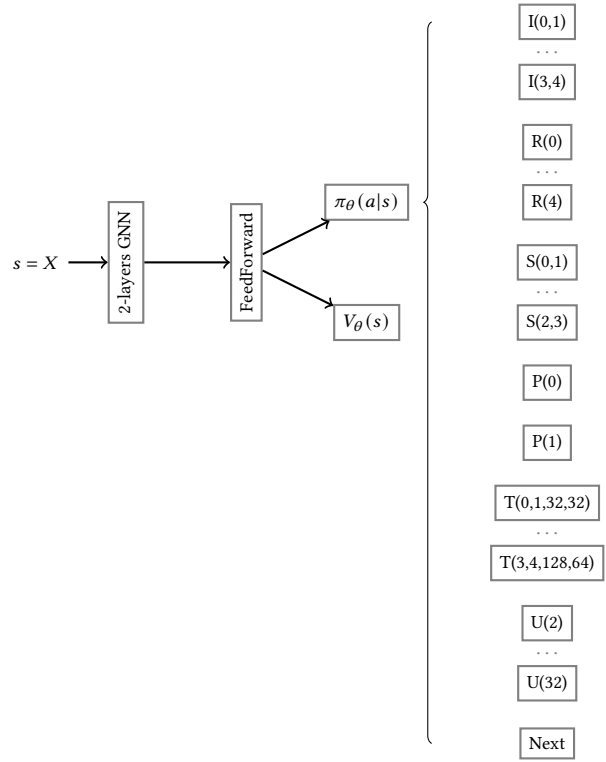
\begin{figure*}[ht]
    \centering
    \tikzset{every node/.style={font=\scriptsize}}  % ensures all fonts are vector-safe
    \begin{tikzpicture}[scale=0.8,
      roundnode/.style={circle, draw=gray, fill=white, thick, minimum size=3mm},
      rectnode/.style={rectangle, draw=gray, fill=white, thick, minimum size=3mm},
      vectornode/.style={rectangle, minimum size=5mm, inner sep=2mm},
    ]  

    \node[rectnode, style={draw=white}] (x) at (-1.5,0) {\textbf{$s=X$}};
    \node[rectnode] (g) at (0,0) {\rotatebox{90}{2-layers GNN}};
    \node[rectnode] (f) at (2,0) {\rotatebox{90}{FeedForward}};
    \node[rectnode] (p) at (4,1) {$\pi_{\theta}(a|s)$};
    \node[rectnode] (v) at (4,-1) {$V_{\theta}(s)$};

  % Vertically aligned action space
    \node[rectnode] (a1) at (7,4) {I(0,1)};
    \node[draw=none] (a2) at (7,3.5) {{\color{gray}\scriptsize$\ldots$}};
    \node[rectnode] (a3) at (7,3) {I(3,4)};
    \node[rectnode] (a4) at (7,2) {R(0)};
    \node[draw=none] (a5) at (7,1.5) {{\color{gray}\scriptsize$\ldots$}};
    \node[rectnode] (a6) at (7,1) {R(4)};
    \node[rectnode] (a7) at (7,0) {S(0,1)};
    \node[draw=none] (a8) at (7,-0.5) {{\color{gray}\scriptsize$\ldots$}};
    \node[rectnode] (a9) at (7,-1) {S(2,3)};
    \node[rectnode] (a10) at (7,-2) {P(0)};
    \node[rectnode] (a11) at (7,-3) {P(1)};
    \node[rectnode] (a12) at (7,-4) {T(0,1,32,32)};
    \node[draw=none] (a13) at (7,-4.5) {{\color{gray}\scriptsize$\ldots$}};
    \node[rectnode] (a14) at (7,-5) {T(3,4,128,64)};
    \node[rectnode] (a15) at (7,-6) {U(2)};
    \node[draw=none] (a16) at (7,-6.5) {{\color{gray}\scriptsize$\ldots$}};
    \node[rectnode] (a17) at (7,-7) {U(32)};
    \node[rectnode] (a18) at (7,-8) {Next};

    % Brace
    \draw[decoration={brace,raise=5pt, aspect=0.75, amplitude=3pt},decorate]
      (5.3,-8) -- node[right=4pt] {} (5.3,4);

    % Model path arrows
    \draw[thick, color=black,->] (x) -- (g); 
    \draw[thick, color=black,->] (g) -- (f); 
    \draw[thick, color=black,->] (f) -- (p); 
    \draw[thick, color=black,->] (f) -- (v); 
    \end{tikzpicture}
    \caption{The action space}
    \label{fig:action-space}
\end{figure*}

\subsubsection{Detailed Actions Space}

The agent's action space consists of 56 actions, each one represents a loop transformation with its parameters. \autoref{fig:action-space} illustrates the output of the policy and the structure of our action space.

\begin{itemize}
    \item I($i,j$) Interchange of loop levels $i$ and $j$ in the targeted branch.
    \item R($i$) Reversal of the loop level $i$ in the targeted branch.
    \item S($i,j$) Skewing of loop levels $i$ and $j$ in the targeted branch.
    \item P($i$) Parallelization of loop level $i$ in the targeted branch.
    \item T($i,j,x,y$) Tiling of loop levels $i$ and $j$ with tile sizes $x$ and $y$ respectively. 
    \item U($x$) Unrolling of the innermost loop level with an unrolling factor $x$.
    \item Next: targets the next branch in the AST.
\end{itemize}

\subsection{Rewards}
The typical performance evaluation in code optimization is defined by the speedup\footnote{Speedup: original execution time divided by the execution time of transformed code.} gained after applying a transformation. The final speedup $\tau_f$ is calculated by multiplying intermediate speedups $\tau_i$, $\tau_f = \displaystyle \prod_{i=1}^n \tau_i$. In reinforcement learning, an agent's goal is to maximize a sum of rewards. To adapt the product of speedups and use it as the agent's reward signal, we use the $log$ function to transform the product into a sum ($\log(\displaystyle\prod_{i=1}^n \tau_i)= \displaystyle\sum_{i=1}^n \log(\tau_i)$) given that the logarithmic function is monotonically increasing. The agent, thus, receives a reward $r_t=\log(a_t)$ instead of $a_t$. Using the logarithm also scales down high values that can increase variance and impact the training's stability.

\subsection{Actor Critic Network} \label{sec:actor-critic}

After extracting the graph-level features $h_{G}^{final}$, we feed it as input to a feed forward network that is divided into a policy to predict the action probabilities, and a state-value approximator. ~\autoref{fig:model} illustrates the overall architecture of the model. The presented architecture was selected based on a series of experimental evaluations (section \ref{Model Architecture Design Choice}).

The GAT layers have 4 attention heads, each with a hidden size of 128. We use a linear function for each layer to project its expanded hidden size multiplied by the number of heads into its original size. After concatenating and getting $h_{G}^{final}$, we pass it through a 2-layer MLP and use the scaled exponential linear unit (SELU) \cite{klambauer2017self} as the activation function. We then separate the network into two heads: a policy head and a state-value head. Both heads have the same architecture, with only the output layer size different. Both are a 3-layer MLP with sizes (128, 128, 56) for the policy head where 56 is the number of actions and (128, 128, 1) for the state-value head. 
\begin{center}
\begin{figure}[H]
\begin{tikzpicture}[
  roundnode/.style={circle, draw=gray, fill=white, thick, minimum size=7mm},
  rectnode/.style={rectangle, draw=gray, fill=white, thick, minimum size=7mm},
  vectornode/.style={rectangle, minimum size=5mm, inner sep=2mm},
  ]  

\node[rectnode, style={draw=white}] (x) at (-1.5,0) {\textbf{$s=X$}};
\node[rectnode, style={draw=white}] (po) at (1,0.3) {\textbf{$h_{G}^{final}$}};
\node[rectnode] (g) at (0,0) {\rotatebox{90}{2-layers GNN}};
\node[rectnode] (f) at (2,0) {\rotatebox{90}{FeedForward}};
\node[rectnode] (p) at (4,1) {$\pi_{\theta}(a|s)$};
\node[rectnode] (v) at (4,-1) {$V_{\theta}(s)$};

\draw[thick, color=black,->] (x) -- (g); 
\draw[thick, color=black,->] (g) -- (f) ; 
\draw[thick, color=black,->] (f) -- (p); 
\draw[thick, color=black,->] (f) -- (v); 

\end{tikzpicture}
\caption{The architecture of our actor-critic agent. The backbone of this model processes the graph input and produces a vector $h_{G}^{final}$ summarizing the graph's characteristics. Followed by feedforward layers, it is then divided into the policy and value heads.}
\label{fig:model}
\end{figure}
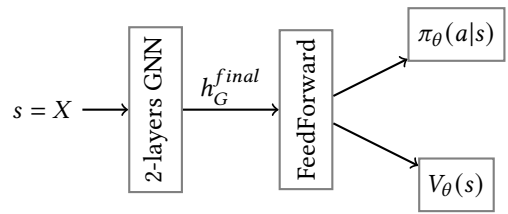
\end{center}

\subsection{Training the Agent}

To train our agent, we use PPO \cite{schulman2017proximal}. When our agent applies an action, the Tiramisu environment tests the legality\footnote{In code optimization, a transformation is deemed legal if its application does not violate the program's data dependencies.} of the action before executing it. We use classical polyhedral dependence analysis and legality checking \cite{feautrier_array_1988,violateddep} to guarantee the correctness of transformations.
As mentioned, we transform speedups using the logarithmic function with a base of 4 to increase the training stability. We assign a speedup of 1 for illegal actions and do not apply them on the original program; we leave it up to the agent to learn what valuable transformations to apply.

\subsection{Training Dataset}

To train the RL agent we used 2,500 randomly generated tiramisu programs. We used the same methodology defined by Baghdadi et al.~\cite{baghdadi2021deep} to generate these programs. The agent explored a considerable number of schedules (sequences of code optimizations) for each program. Overall, the total number of unique schedules that the agent was trained on has reached $\sim$ 45,000.

\section{Accelerating the Training of the RL Agent}

During training, when the RL agent picks an action (a code optimization), it first checks the legality of that action (using classical polyhedral legality check provided in the Tiramisu compiler\cite{baghdadi2019tiramisu}). If the action is legal, the agent then computes its reward. To do that, it applies the code optimization and compiles the optimized program and runs it on the target hardware and then computes the execution time of the optimized program and use that to compute the reward. These three steps: checking legality, compiling code and running code on the target hardware are time consuming. For example, checking the legality of code optimizations takes about 30\% of the training time in our RL system, while compiling and running code takes the majority of the remaining time. Because these steps are time consuming, a single training of the RL agent takes weeks. During the development phase of the RL system, one needs to train the RL system many times, to test different features, ideas, and to fine-tune the hyper-parameters. Since every training takes weeks, the development of the RL system becomes challenging. In this section, we propose methods to mitigate this issue.

\subsection{Execution Time and Legality Check Memoization}

The main idea for this method is to store the result of the legality check and execution times for the actions explored by the agent during a given training, and reuse them in future trainings. 

We construct a dataset, where we store the result of the legality check and execution time obtained for each action chosen by the RL agent. The dataset has the following components:

\begin{enumerate}
    \item \textbf{Programs:} the list of all the randomly generated programs used to train the RL agent.
    \item \textbf{Schedules:} the list of schedules explored by the RL agent for each program (a schedule is a sequence of code optimizations).
    \item \textbf{Schedule legality:} for each pair of (program, schedule) in the dataset, we store the legality of the schedule, as a boolean value.
    \item \textbf{Execution times:} for each pair of (program, schedule) in the dataset, we store the execution time of the \emph{program} when optimized using the \emph{schedule}. The reward can be easily derived from this execution time.
\end{enumerate}

In subsequent trainings, when the RL agent picks an action, it first queries the dataset to check if the corresponding program and schedule exist. If found, the legality check and execution times stored in the dataset are retrieved, bypassing the need for the legality check, compilation, and execution. This significantly reduces redundant computations. If the legality check and execution times are not found, the agent performs the legality check, compiles and executes the optimized program and updates the dataset, enriching it for future use. Initially, when the dataset is empty, the training is slow. Once an initial training is performed, future trainings are likely to be faster. This is particularly true because RL agents, when they converge, tend to pick the same actions, and therefore the probability of picking an action that has been explored in previous trainings is high.

\subsection{Actor-Critic Pre-training}
\label{Actor-Critic Pre-training}

The goal of this technique is to enable better learning in the actor-critic neural networks. That would translate in either faster convergence, or to a convergence to a higher average reward. We improve the learning in the actor-critic neural networks by initializing the weights of the actor-critic neural networks through actor-critic pre-training (i.e., by pre-training the actor-critic neural networks).

The GNN layers, the feed forward, and the value network {$V_{\theta}(s)$} (mentioned in ~\autoref{fig:model}) are trained, before training the RL agent, on a surrogate task: predicting the execution times of unoptimized programs from their graph representation. The weight of these layers are then used as an initialization for the corresponding layers in the RL agent. Predicting the execution times of programs aligns well with the goals of pre-training the actor-critic agent because it is a complex task that provides a substantial amount of knowledge about the differences between programs. While a more targeted pre-training, such as using data in the format \emph{(program, schedule, speedup)}, could potentially be better, the lack of sufficient data and the significant time required to generate a large dataset directed our choice toward execution time prediction. Our goal was not to create a highly accurate execution time prediction model but to provide the agent with reasonable initial weights. We believe that the current approach is sufficient to achieve that. We plan to extend our work in the future to pre-train on data of the format  \emph{(program, schedule, speedup)}.

Note that even if we train the RL agent with higher quality data, we do not have guarantees on faster convergence. This is mainly because even if we pre-train the actor-critic with such data, the RL agent would still need to explore actions initially with a certain degree of randomness, due to the use the epsilon greedy algorithm~\cite{sutton2018reinforcement} in training the RL agent, a common method for balancing exploration and exploitation in training reinforcement learning agents. In epsilon greedy, an entropy coefficient is set to a high value initially and then decays over time to reach zero. Higher values of the entropy coefficient force the RL agent to explore actions more initially, and over time during training, the entropy becomes smaller, allowing the RL agent to converge. Because the agent has to explore actions with a degree of randomness in its initial phase of training (while the entropy coefficient is high), it will continue exploring (with a degree of randomness) even if it converges earlier, making the training last for longer.

We pre-trained the layers (GNN layers, feedforward layers and the value network) on 26,000 data points. The model input is the graph representation of a program, and the output is the execution time of the program (unoptimized program). We use the MSE (Mean Squared Error) Loss and train for 1500 epochs with a $10^{-4}$ learning rate. The resulting weights are then used to initialize the GNN layers, feedforward layers and the value network layers.

\section{Experiments and Evaluation}

\subsection{Experimental Setup}

To train our reinforcement learning agent, we used a cluster where each node is a 28-core Intel(R) Xeon(R) CPU E5-2680 v4 @ 2.40GHz, $4$ GB of RAM per core. The OS installed on the nodes is CentOS Linux version $8$. We used distributed learning to train the model using 4 nodes of the cluster in all the upcoming evaluations. 

\subsection{Training Details}

To train the GNN with PPO we used the following set of hyperparameters as specified in \autoref{tab:hyperparams}.

\begin{table}[h]
\caption{Parameters of training the agent.}
\label{tab:hyperparams}
\vspace{-0.4cm}
\begin{center}
\begin{small}
%\begin{sc}
\begin{tabular}{|l@{\hspace{0.2cm}}|c@{\hspace{0.3cm}}|c|c|r|}

\hline
Parameters (PPO) & Value \\ 
\hline
$\epsilon_{clip}$ & 0.3 \\
$\gamma$ &  0.99 \\
$\lambda$ & 0.95\\
Value coefficient & 2 \\
Entropy coefficient (decaying) & $10^{-1} \rightarrow 0$ \\
Batch size & 512 \\
Num epochs & 5 \\
Mini-Batch size & 64 \\
Learning rate & $10^{-4}$\\
\hline
Parameters (GNN) &  \\ 
\hline
Type & GATv2 \cite{brody2022attentive}\\
Num layers & 2 \\
Num of attention heads & 4 \\
Hidden layers size & 128 \\
\hline
\end{tabular}
%\end{sc}
\end{small}
\end{center}
\end{table}

\subsection{Evaluation on a Benchmark Suite}
\label{Evaluation on a Benchmark Suite}

In this section, we evaluate how effective is our RL agent in optimizing real-world benchmarks. For this evaluation, we use the same benchmark suite used by \citet{baghdadi2021deep}, a set of benchmarks including image processing, deep learning, and linear algebra programs. For each benchmark, we execute the schedule obtained by the RL agent 30 times before taking the minimum to reduce the effect of noise on the results. 
The following are descriptions of the benchmarks used:

\begin{itemize}
    \item \textbf{blur:} an image processing filter to blur images.
    \item \textbf{cvtcolor:} an image processing filter for converting the colors of an input image from RGB to grayscale.
    \item \textbf{doitgen:} a kernel from the multiresolution adaptive numerical scientific simulation \cite{pouchet2012polybench}.
    \item \textbf{heat2d: } Heat equation over 2D data domain.
    \item \textbf{heat3d: } Heat equation over 3D data domain.
    \item \textbf{jacobi2d: } a jacobi-style stencil computation over 2D data with 5-point stencil pattern.
    \item \textbf{mvt: } matrix-vector multiplication composed with another matrix-vector multiplication but with a transposed matrix.
    \item \textbf{seidel-2d:} two dimensional Seidel stencil computation.
\end{itemize}

We compare the speedups we get using our proposed system to those produced by the Tiramisu autoscheduler~\cite{baghdadi2021deep} (Tiramisu's search-based automatic code optimization). The speedup of an optimized program is defined as follows:
\begin{multline}
speedup = \frac{exec\_time\_unoptimized\_program}{exec\_time\_optimized\_program}
\end{multline}
Speedups higher than 1 indicate that the optimized program is faster than the original one. The baseline of computing the speedups in these experiments is the execution time of the unoptimized program.

We also compare the speedups of our proposed system to Pluto, a state-of-the-art polyhedral compiler that does not use machine learning (it uses Integer Linear Programming). Pluto, being a polyhedral compiler like Tiramisu, supports a large space of complex code transformations (we used Pluto with the \textit{--parallel} \textit{--tile} options to enable parallelism and tiling). We also compare to HalideRL, a state-of-the-art compiler that uses reinforcement learning for automatic code optimization, and that we consider to be the closest to our work. HalideRL trains the RL on each one of the benchmarks. We perform the training on our target machine (i.e., on our cluster nodes) and leave it until it converges. We do not compare with PolyGym in this experiment because PolyGym focuses mainly on proposing an RL environment and does not propose a deep RL agent for the environment.

The performance of our proposed system, compared to the state-of-the-art, is presented in ~\autoref{tab:benchmark_eval}.
Our agent predicts code optimizations that lead to a geometric mean speedup reaching $3.16\times$ over unoptimized code. It also has a geometric mean speedup higher than the Tiramisu autoscheduler, Pluto and HalideRL, with a geometric mean speedup of \speedupOverTiramisu over the Tiramisu autoscheduler. HalideRL crashed for 3 benchmarks (\emph{blur}, \emph{doitgen}, and \emph{jacobi2d}), and therefore we did not report speedups for those.

Our system is the only one where the agent found only useful code optimizations. In other words, it did not choose optimizations that led to a slowdown compared to the original unoptimized program (all the speedups are $>= 1$).

\begin{table}[h]
\caption{Summary of results regarding execution time speedup achieved by each method. The baseline of computing the speedups is the original execution time of the functions without any transformation applied.}
\label{tab:benchmark_eval}
% \vskip 0.15in
\begin{center}
%\begin{small}
%\begin{sc}
\begin{tabular}{|l@{\hspace{0.2cm}}|c@{\hspace{0.3cm}}|c|c|r|}

\hline
Benchmark & Tiramisu & Pluto  & HalideRL & \toolName{} \\ 
 & Autoscheduler & & &  \\
 
\hline

blur & 0.27 &  1.01 & / & \textbf{4.27}\\
cvtcolor & \textbf{1.12} & 0.90& 0.14 &  1 \\
doitgen & 2.66 & 0.74 & / & \textbf{11.37} \\
heat2d & 1.86 & 0.98  & 1.15 & \textbf{2.39}\\
heat3d & 0.82 & 1.01 & $3.10^{-3}$ &  \textbf{2.36}\\
jacobi2d & \textbf{1.66} & 1 & / & 1\\
mvt & 4.14 &  0.97& 0.27 & \textbf{6.1} \\
seidel2d & 4.24 & 0.99 & 5.57  & \textbf{6.03}\\
\hline
geo mean     & 1.56  & 0.94  & 0.23 & \textbf{3.16} \\
\hline

\end{tabular}
%\end{sc}
%\end{small}
\end{center}
\end{table}

High speedups in benchmarks such as \emph{doitgen}, \emph{mvt}, \emph{seidel2d}, \emph{blur}, \emph{heat2d}, and \emph{heat3d} are due to the application of parallelization and tiling which improves data locality. The agent refrained from parallelizing code in cases where parallelization leads to a decrease in performance (if the overhead of parallelization is higher than its benefit). This was the case for \emph{cvtcolor}, for example, where the outer loop is the color channel and has only 3 iterations. Parallelizing such a loop leads to high overhead with little benefit. The Tiramisu autoscheduler could obtain better speedups than our agent in this case because it applied another transformation (loop interchange) that interchanged one of the inner loops (which represents the image height) to become the outer loop and then parallelized that loop. Since the new outer loop has a high number of iterations, parallelization was beneficial and therefore the Tiramisu autoscheduler obtained a higher speedup.

In summary, the evaluation shows that our proposed agent was able to outperform three state-of-the-art compilers. The first uses a search-based method (Tiramisu autoscheduler), the second uses integer linear programming (Pluto), while the third uses RL (HalideRL). Among these three, the Tiramisu autoscheduler achieved the best speedups, but our proposed RL ouperformed the Tiramisu autoscheduler by \speedupOverTiramisu, highlighting the benefit of an RL-based approach.

\begin{figure}[H]
    \vspace{-0.4cm}
    \centering
    \includegraphics[width=\columnwidth]{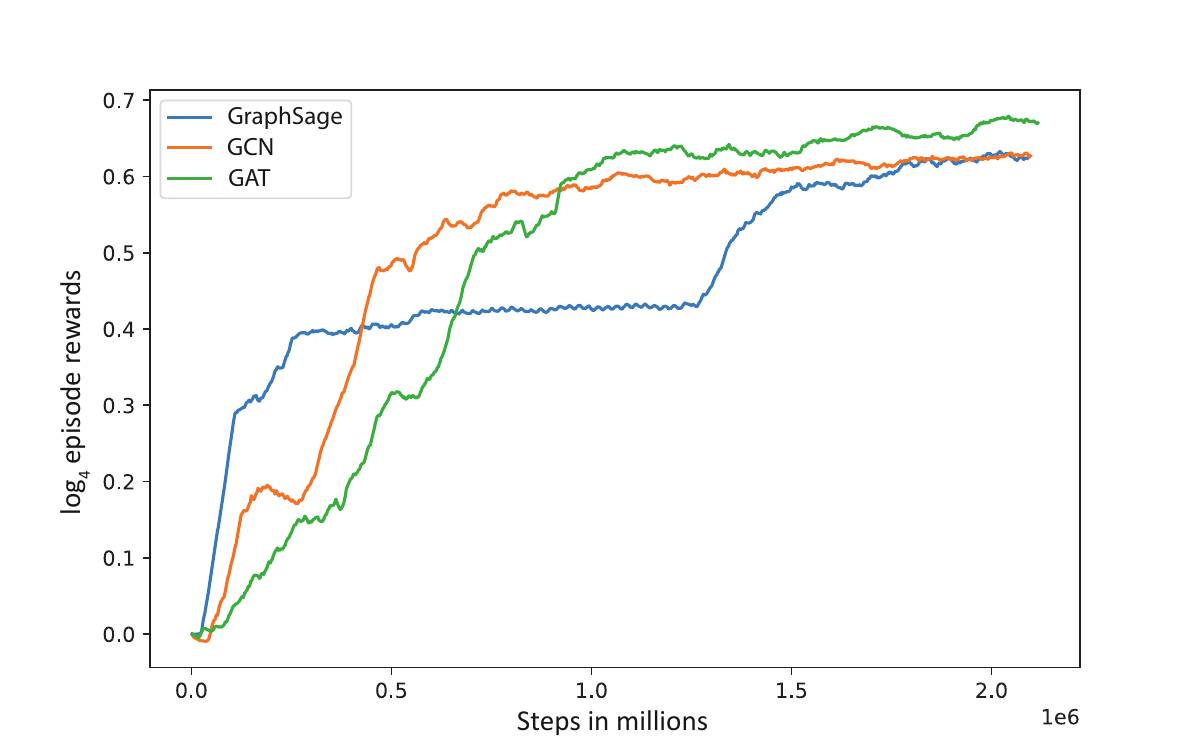}
    \vspace{-0.5cm}
    \caption{The performance of three agents using different GNN layers, the y-axis represents the average $log_4$ of the episodes speedups collected in a single PPO iteration. The x-axis represents the number of actions taken in total.}
    \label{fig:ablation}

    \begin{figure}[H]
    \centering
    \includegraphics[width=1.02\columnwidth]{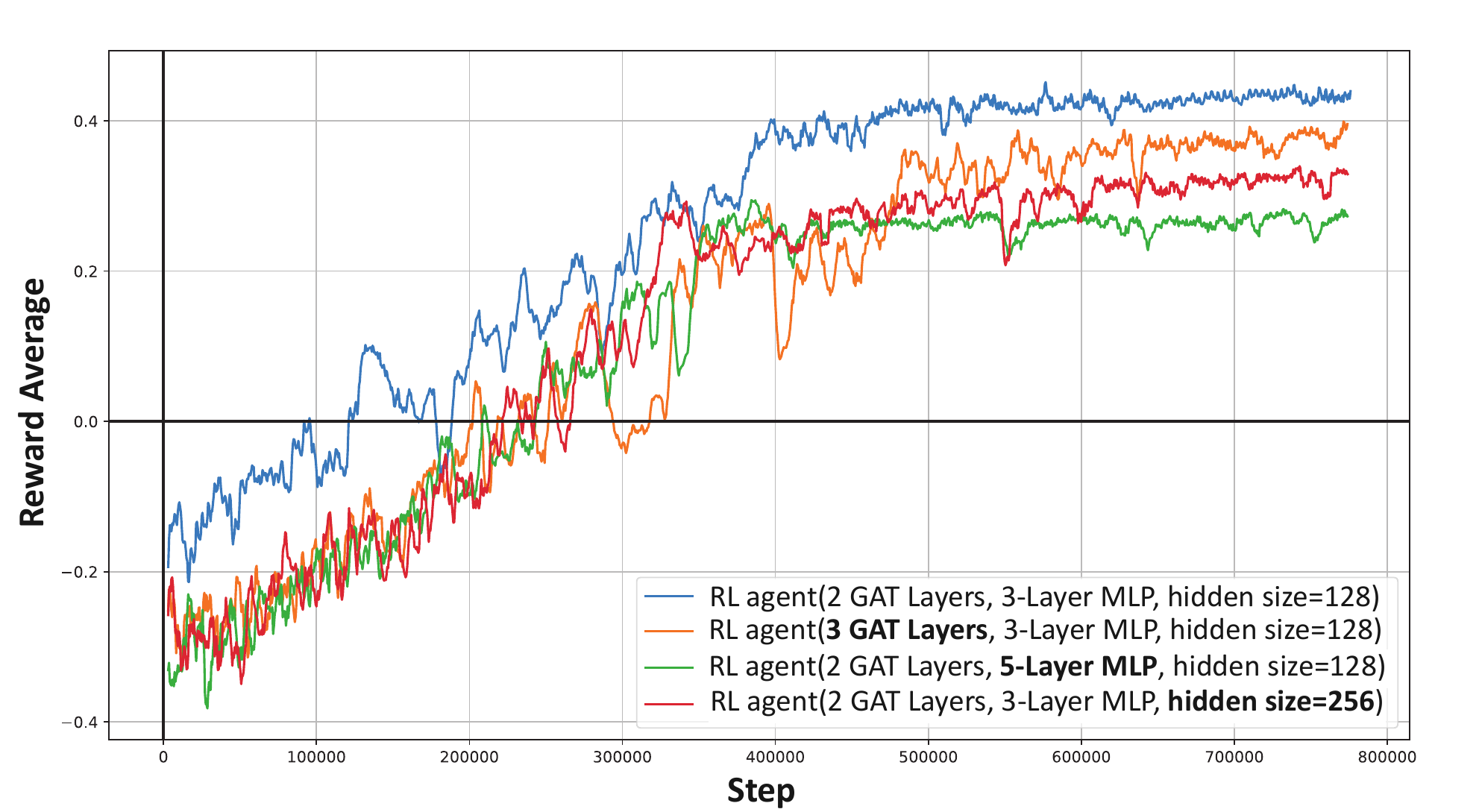} % Replace with your image
    \caption{Reward Averages Across Different Experiments over Training Steps}
    \label{fig:archiChoices}
\end{figure}

\end{figure}
\subsection{Model Architecture Design Choices \label{Model Architecture Design Choice}}

\subsubsection{Evaluating Different GNN Models}
Alongside the GAT architecture that we use in this work, we evaluated other prominent GNN models such as Graph Convolutional Networks (GCN) \cite{kipf2016semi} and GraphSAGE \cite{hamilton2017inductive}. Figure~\ref{fig:ablation} depicts the comparison between training the three agents with PPO.

For the three agents, we use the same architecture described in \ref{sec:actor-critic} except for the GNN layer type we want to test. We notice a slightly better performance of the GAT agent over GCN and GraphSage. Note that we did this experiment early in the lifetime of the project, on a subset of our training dataset. We believe that the results would generalize to the whole dataset though.

\subsubsection{Number of GAT and MLP Layers}
In this section, we evaluate different variants of our proposed model.
Our focus in these experiments was on the number of layers of GATs, MLPs, and the size of the hidden layers, as these affect the ability of the model to learn complex patterns and relations in the data. The best-performing agent uses 2 GAT layers, 3 MLP layers and a size of the hidden layers equal to 128. We tried out different other configurations that presented in  \autoref{fig:archiChoices}.

\subsection{Evaluating the Execution Time and Legality Check Memoization Method}

\begin{center}
\begin{figure}[H]
\includegraphics[width=1.02\columnwidth]{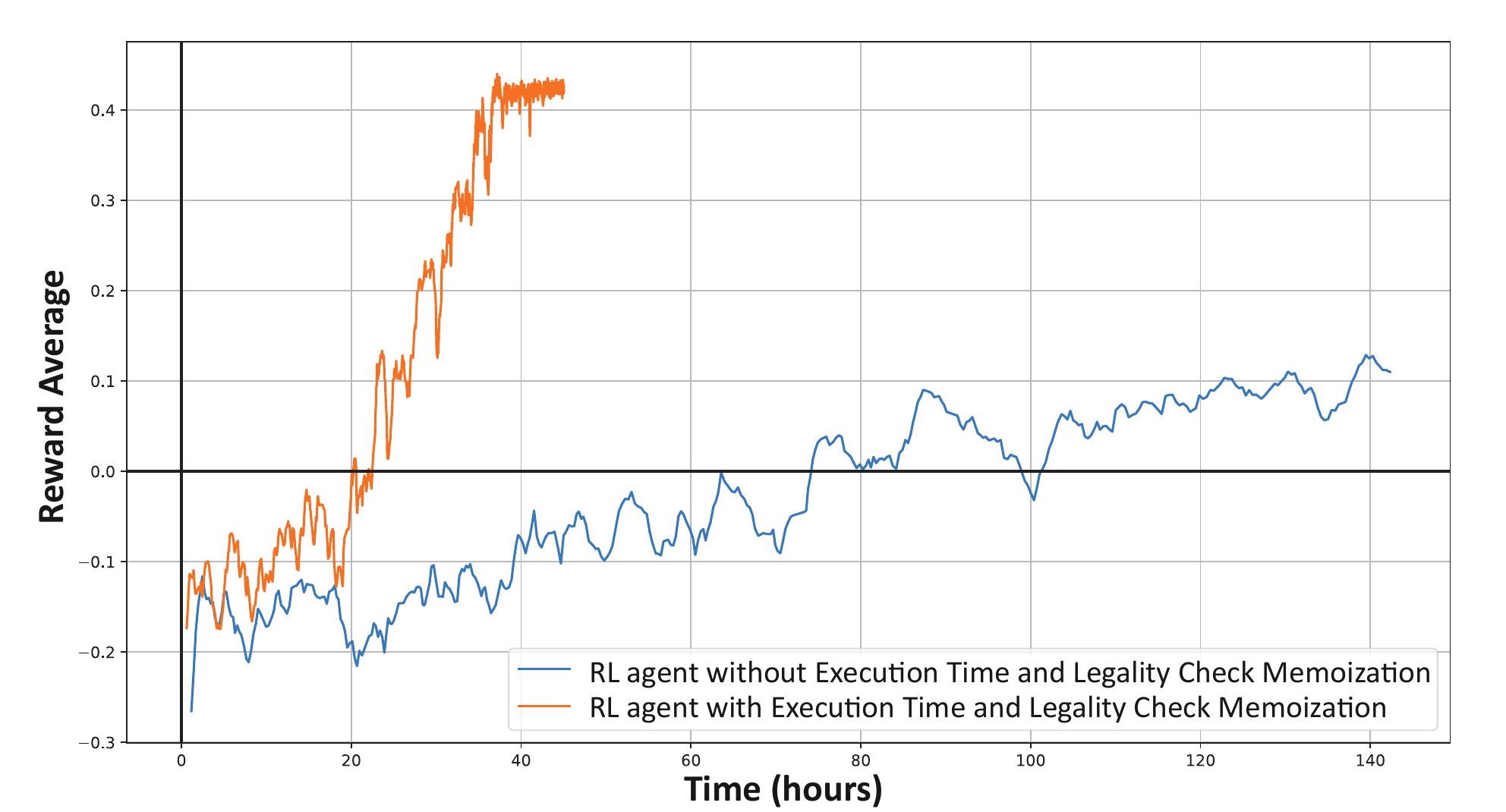}
\caption{Reward Averages of our Agents With and Without Execution Time and Legality Check Memoization}
\label{fig:caching}
\end{figure}
\end{center}

To analyze the efficiency of the memoization technique, we evaluated the training time of our proposed RL agent with and without memoization.
\autoref{fig:caching} shows a significant reduction in convergence time for the agent that uses memoization. It converges to the best average reward in 45 hours.

%This accelerates the training process of our RL agents, enabling more extensive experimentation and improved scalability.

\begin{center}
\begin{figure}[H]
\includegraphics[width=0.9\columnwidth]{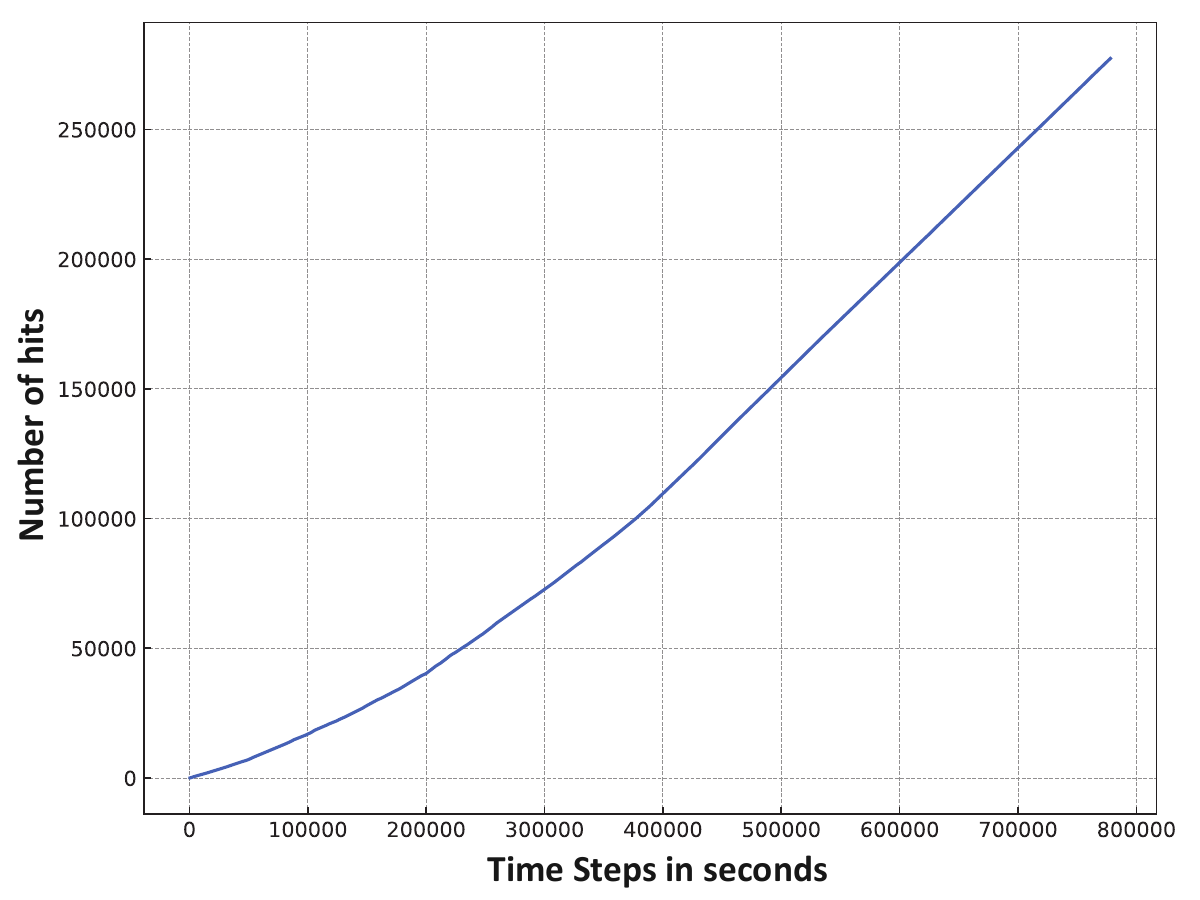}

\vspace{-0.25cm}
\caption{Number of hits during the training of the RL agent.}
\label{fig:numHits}
\vspace{-0.5cm}
\end{figure}
\end{center}

To further evaluate the technique of memoization, we recorded the total number of hits while training our RL agent. A hit in this case indicates that a schedule (with its legality and execution time) is already present in the dataset. \autoref{fig:numHits} shows how the total number of hits increases as the training progresses, which in turn indicates that the stored values are indeed being used during the RL training.
\vspace{1em} 
\vspace{-3em} 
\subsection{Evaluating the Actor-Critic Pre-training}

To evaluate the actor-critic pre-training technique, we trained two agents: the first, without pre-training, while the second uses the pre-training technique mentioned in Section \ref{Actor-Critic Pre-training}. In these experiments, we use the memoization technique as we have already demonstrated its effectiveness. We recorded the average reward during training, the training time, and then evaluated the trained RL agents on the benchmark suite used in section \ref{Evaluation on a Benchmark Suite}.

\begin{center}
    \begin{figure}[ht]
        \includegraphics[width=\columnwidth]{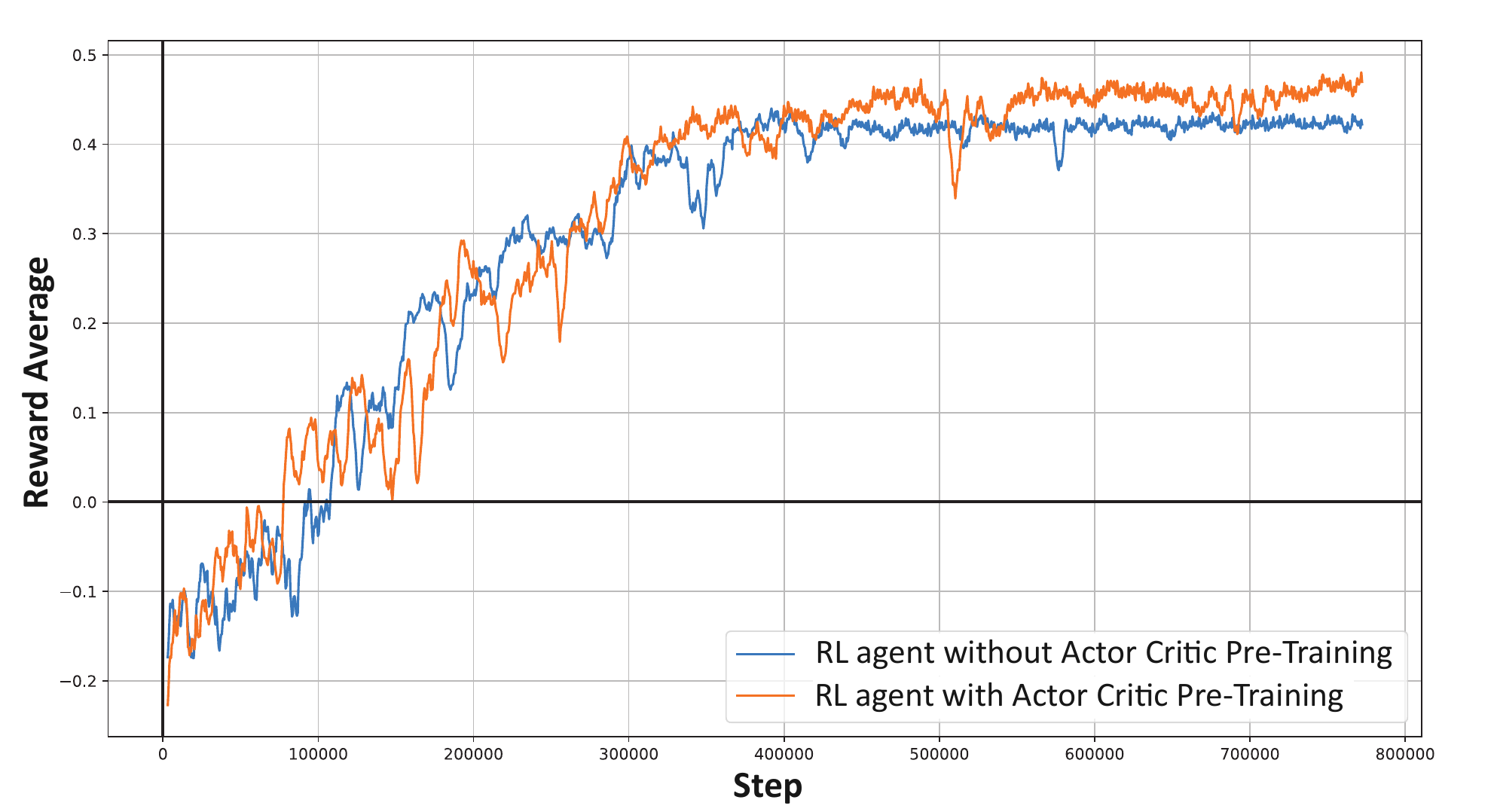} % Replace with your image
        \vspace{-0.5cm}
        \caption{Reward Averages of our Agents With and Without Actor-Critic Pre-training}
        \label{fig:pretraining}
        \end{figure}
    \end{center}

\autoref{fig:pretraining} shows the average reward of the two RL agents. The RL agent that used pre-training managed to obtain a much higher average reward, compared to the agent that did not. This difference in the average reward is significant though, since the reward is the \emph{log} of the speedup, and therefore a small difference in the average reward translates to much larger difference in speedups. \autoref{tab:pretraining_eval} shows an evaluation of the two RL agents on the benchmark. The RL that used the pre-training method obtained significantly better speedups on the benchmarks with nearly the same training times (\autoref{fig:pretraining}). The agent without pre-training failed to optimize the \emph{Heat2d} benchmark. In the case of \emph{seidel2d} it unrolled the second loop instead of parallelizing the outermost loop and then applying tiling. We clearly observe that this agent did not learn the importance of parallelizing outermost loops, among other patterns.

In this experiment, we did not notice faster convergence of the RL agent because of the high initial entropy used in this training. This high entropy is important for the agent to learn useful code optimizations though, and lower entropy values would lead to a lower average reward. The actor-critic pre-training method was useful in allowing the actor-critic neural networks to better learn though, and this translates in obtaining a better average reward.

\begin{table}[h]
\caption{Speedups achieved by Pearl with and without Actor-Critic Pre-training}
\label{tab:pretraining_eval}
\vspace{-0.25cm}
\begin{center}
%\begin{small}
%\begin{sc}
\begin{tabular}{|l@{\hspace{0.2cm}}|c@{\hspace{0.3cm}}|c@{\hspace{0.3cm}}|}

\hline
Benchmark & Pearl  & Pearl  \\ 
 & with Pre-Training &  without Pre-Training  \\
 
\hline

blur & \textbf{4.27} &  4.17 \\
cvtcolor & \textbf{1}  & 1\\
doitgen & 11.37 & \textbf{12.09} \\
heat2d & \textbf{2.39} & 1  \\
heat3d & 2.36 & \textbf{2.39} \\
jacobi2d & \textbf{1} & 1\\
mvt & \textbf{6.1} &  5.98 \\
seidel2d & \textbf{6.03} & 0.65 \\
\hline
geo mean     & \textbf{3.16}  & 2.15   \\
\hline

\end{tabular}
\vspace{-0.5cm}
%\end{sc}
%\end{small}
\end{center}
\end{table}

\subsection{Search Space Exploration Time Comparison}
Our RL system not only has better performance compared to the existing Tiramisu auto-scheduler, but it also does so significantly faster, which is one of the key contributions of this work. Reinforcement learning has the advantage of learning a policy network (actor) that directly predicts the sequence of code optimizations to apply to obtain the best performance. This is in contrast to current state-of-the-art methods that use tree-search methods to explore the search space using a tree-search method (e.g., beam-search). Tree-search methods are time-consuming compared to a policy network learned through reinforcement learning. On the reported benchmarks, the RL system identified the sequence of cost optimizations to use in \emph{33.36 milliseconds} on average, which is $563.67\times$ faster than the Tiramisu scheduler~\cite{baghdadi2021deep}.

\section{Discussion and Future Work}
 
In its current state, our proposed system supports six types of loop transformations. These transformations, along with their parameters and the choices of which loops they apply to, constitute a large search space reaching $10^{170}$ candidates~\cite{adams2019halide,baghdadi2021deep}. Although this search space is already large, we plan to support a wider set of loops and data layout transformations.
We also plan to explore the effect of applying RL on raw text data instead of encoding the states.

\section{Conclusion}

This paper introduces a deep-reinforcement learning-based autoscheduler for the Tiramisu compiler. We train an RL agent on a dataset of synthetic Tiramisu programs.

During training, the policy network of the agent converges into a heuristic that we use to infer schedules for unseen programs. By evaluating our proposed system on standard benchmarks, we show its competitiveness with state-of-the-art autoschedulers. Compared to Tiramisu, our RL-based agent achieves an overall geometric mean speedup of \speedupOverTiramisu.

\begin{acks}

This research has been partly supported by the Center for Artificial Intelligence and Robotics (CAIR) at New York University Abu Dhabi, funded by Tamkeen under the NYUAD Research Institute Award CG010. The research was carried out on the High-Performance Computing resources at New York University Abu Dhabi.
\end{acks}

\bibliographystyle{ACM-Reference-Format}
\bibliography{references}

%%% -*-BibTeX-*-
%%% Do NOT edit. File created by BibTeX with style
%%% ACM-Reference-Format-Journals [18-Jan-2012].

\begin{thebibliography}{59}

%%% ====================================================================
%%% NOTE TO THE USER: you can override these defaults by providing
%%% customized versions of any of these macros before the \bibliography
%%% command.  Each of them MUST provide its own final punctuation,
%%% except for \shownote{}, \showDOI{}, and \showURL{}.  The latter two
%%% do not use final punctuation, in order to avoid confusing it with
%%% the Web address.
%%%
%%% To suppress output of a particular field, define its macro to expand
%%% to an empty string, or better, \unskip, like this:
%%%
%%% \newcommand{\showDOI}[1]{\unskip}   % LaTeX syntax
%%%
%%% \def \showDOI #1{\unskip}           % plain TeX syntax
%%%
%%% ====================================================================

\ifx \showCODEN    \undefined \def \showCODEN     #1{\unskip}     \fi
\ifx \showDOI      \undefined \def \showDOI       #1{#1}\fi
\ifx \showISBNx    \undefined \def \showISBNx     #1{\unskip}     \fi
\ifx \showISBNxiii \undefined \def \showISBNxiii  #1{\unskip}     \fi
\ifx \showISSN     \undefined \def \showISSN      #1{\unskip}     \fi
\ifx \showLCCN     \undefined \def \showLCCN      #1{\unskip}     \fi
\ifx \shownote     \undefined \def \shownote      #1{#1}          \fi
\ifx \showarticletitle \undefined \def \showarticletitle #1{#1}   \fi
\ifx \showURL      \undefined \def \showURL       {\relax}        \fi
% The following commands are used for tagged output and should be
% invisible to TeX
\providecommand\bibfield[2]{#2}
\providecommand\bibinfo[2]{#2}
\providecommand\natexlab[1]{#1}
\providecommand\showeprint[2][]{arXiv:#2}

\bibitem[Adams et~al\mbox{.}(2019)]%
        {adams2019halide}
\bibfield{author}{\bibinfo{person}{Andrew Adams}, \bibinfo{person}{Karima Ma}, \bibinfo{person}{Luke Anderson}, \bibinfo{person}{Riyadh Baghdadi}, \bibinfo{person}{Tzu-Mao Li}, \bibinfo{person}{Micha\"{e}l Gharbi}, \bibinfo{person}{Benoit Steiner}, \bibinfo{person}{Steven Johnson}, \bibinfo{person}{Kayvon Fatahalian}, \bibinfo{person}{Fr\'{e}do Durand}, {and} \bibinfo{person}{Jonathan Ragan-Kelley}.} \bibinfo{year}{2019}\natexlab{}.
\newblock \showarticletitle{Learning to Optimize Halide with Tree Search and Random Programs}.
\newblock \bibinfo{journal}{\emph{ACM Trans. Graph.}} \bibinfo{volume}{38}, \bibinfo{number}{4}, Article \bibinfo{articleno}{121} (\bibinfo{date}{jul} \bibinfo{year}{2019}), \bibinfo{numpages}{12}~pages.
\newblock
\showISSN{0730-0301}
\urldef\tempurl%
\url{https://doi.org/10.1145/3306346.3322967}
\showDOI{\tempurl}


\bibitem[Ahn et~al\mbox{.}(2020)]%
        {ahn2020chameleon}
\bibfield{author}{\bibinfo{person}{Byung~Hoon Ahn}, \bibinfo{person}{Prannoy Pilligundla}, \bibinfo{person}{Amir Yazdanbakhsh}, {and} \bibinfo{person}{Hadi Esmaeilzadeh}.} \bibinfo{year}{2020}\natexlab{}.
\newblock \bibinfo{title}{Chameleon: Adaptive Code Optimization for Expedited Deep Neural Network Compilation}.
\newblock
\newblock
\showeprint[arxiv]{2001.08743}


\bibitem[Baghdadi(2015)]%
        {baghdadi2015PhD}
\bibfield{author}{\bibinfo{person}{Mohamed~Riyadh Baghdadi}.} \bibinfo{year}{2015}\natexlab{}.
\newblock \emph{\bibinfo{title}{Improving tiling, reducing compilation time, and extending the scope of polyhedral compilation}}.
\newblock \bibinfo{thesistype}{Ph.\,D. Dissertation}. \bibinfo{school}{Paris 6}.
\newblock


\bibitem[Baghdadi et~al\mbox{.}(2011)]%
        {baghdadi2011speculation}
\bibfield{author}{\bibinfo{person}{Riyadh Baghdadi}, \bibinfo{person}{Albert Cohen}, \bibinfo{person}{Cedric Bastoul}, \bibinfo{person}{Louis-Noel Pouchet}, {and} \bibinfo{person}{Lawrence Rauchwerger}.} \bibinfo{year}{2011}\natexlab{}.
\newblock \bibinfo{title}{The Potential of Synergistic Static, Dynamic and Speculative Loop Nest Optimizations for Automatic Parallelization}.
\newblock
\newblock
\showeprint[arxiv]{1111.6756}~[cs.DC]


\bibitem[Baghdadi et~al\mbox{.}(2015)]%
        {baghdadi2015pencil}
\bibfield{author}{\bibinfo{person}{Riyadh Baghdadi}, \bibinfo{person}{Albert Cohen}, \bibinfo{person}{Tobias Grosser}, \bibinfo{person}{Sven Verdoolaege}, \bibinfo{person}{Javed Absar}, \bibinfo{person}{Sven Van~Haastregt}, \bibinfo{person}{Alexey Kravets}, \bibinfo{person}{Anton Lokhmotov}, {and} \bibinfo{person}{Alastair Donaldson}.} \bibinfo{year}{2015}\natexlab{}.
\newblock \emph{\bibinfo{title}{PENCIL Language Specification}}.
\newblock \bibinfo{thesistype}{Ph.\,D. Dissertation}. \bibinfo{school}{INRIA}.
\newblock


\bibitem[Baghdadi et~al\mbox{.}(2013)]%
        {baghdadi2013pencil}
\bibfield{author}{\bibinfo{person}{Riyadh Baghdadi}, \bibinfo{person}{Albert Cohen}, \bibinfo{person}{Serge Guelton}, \bibinfo{person}{Sven Verdoolaege}, \bibinfo{person}{Jun Inoue}, \bibinfo{person}{Tobias Grosser}, \bibinfo{person}{Georgia Kouveli}, \bibinfo{person}{Alexey Kravets}, \bibinfo{person}{Anton Lokhmotov}, \bibinfo{person}{Cedric Nugteren}, {et~al\mbox{.}}} \bibinfo{year}{2013}\natexlab{}.
\newblock \showarticletitle{PENCIL: Towards a platform-neutral compute intermediate language for DSLs}.
\newblock \bibinfo{journal}{\emph{arXiv preprint arXiv:1302.5586}} (\bibinfo{year}{2013}).
\newblock


\bibitem[Baghdadi et~al\mbox{.}(2020)]%
        {baghdadi2020tiramisuDNNDenseSparse}
\bibfield{author}{\bibinfo{person}{Riyadh Baghdadi}, \bibinfo{person}{Abdelkader~Nadir Debbagh}, \bibinfo{person}{Kamel Abdous}, \bibinfo{person}{Fatima~Zohra Benhamida}, \bibinfo{person}{Alex Renda}, \bibinfo{person}{Jonathan~Elliott Frankle}, \bibinfo{person}{Michael Carbin}, {and} \bibinfo{person}{Saman Amarasinghe}.} \bibinfo{year}{2020}\natexlab{}.
\newblock \bibinfo{title}{TIRAMISU: A Polyhedral Compiler for Dense and Sparse Deep Learning}.
\newblock
\newblock
\showeprint[arxiv]{2005.04091}~[cs.DC]


\bibitem[Baghdadi et~al\mbox{.}(2021)]%
        {baghdadi2021deep}
\bibfield{author}{\bibinfo{person}{Riyadh Baghdadi}, \bibinfo{person}{Massinissa Merouani}, \bibinfo{person}{Mohamed-Hicham Leghettas}, \bibinfo{person}{Kamel Abdous}, \bibinfo{person}{Taha Arbaoui}, \bibinfo{person}{Karima Benatchba}, {et~al\mbox{.}}} \bibinfo{year}{2021}\natexlab{}.
\newblock \showarticletitle{A deep learning based cost model for automatic code optimization}.
\newblock \bibinfo{journal}{\emph{Proceedings of Machine Learning and Systems}}  \bibinfo{volume}{3} (\bibinfo{year}{2021}), \bibinfo{pages}{181--193}.
\newblock


\bibitem[Baghdadi et~al\mbox{.}(2019)]%
        {baghdadi2019tiramisu}
\bibfield{author}{\bibinfo{person}{Riyadh Baghdadi}, \bibinfo{person}{Jessica Ray}, \bibinfo{person}{Malek~Ben Romdhane}, \bibinfo{person}{Emanuele Del~Sozzo}, \bibinfo{person}{Abdurrahman Akkas}, \bibinfo{person}{Yunming Zhang}, \bibinfo{person}{Patricia Suriana}, \bibinfo{person}{Shoaib Kamil}, {and} \bibinfo{person}{Saman Amarasinghe}.} \bibinfo{year}{2019}\natexlab{}.
\newblock \showarticletitle{Tiramisu: A polyhedral compiler for expressing fast and portable code}. In \bibinfo{booktitle}{\emph{2019 IEEE/ACM International Symposium on Code Generation and Optimization (CGO)}}. IEEE, \bibinfo{pages}{193--205}.
\newblock


\bibitem[Baghdadi et~al\mbox{.}(2018)]%
        {baghdadi2018tiramisu1}
\bibfield{author}{\bibinfo{person}{Riyadh Baghdadi}, \bibinfo{person}{Jessica Ray}, \bibinfo{person}{Malek~Ben Romdhane}, \bibinfo{person}{Emanuele Del~Sozzo}, \bibinfo{person}{Patricia Suriana}, \bibinfo{person}{Shoaib Kamil}, {and} \bibinfo{person}{Saman~P Amarasinghe}.} \bibinfo{year}{2018}\natexlab{}.
\newblock \showarticletitle{Tiramisu: A code optimization framework for high performance systems}.
\newblock \bibinfo{journal}{\emph{arXiv preprint arXiv:1804.10694}} (\bibinfo{year}{2018}).
\newblock


\bibitem[Bondhugula et~al\mbox{.}(2008a)]%
        {bondhugula2008pluto}
\bibfield{author}{\bibinfo{person}{Uday Bondhugula}, \bibinfo{person}{Albert Hartono}, \bibinfo{person}{J Ramanujam}, {and} \bibinfo{person}{P Sadayappan}.} \bibinfo{year}{2008}\natexlab{a}.
\newblock \showarticletitle{Pluto: A practical and fully automatic polyhedral program optimization system}. In \bibinfo{booktitle}{\emph{Proceedings of the ACM SIGPLAN 2008 Conference on Programming Language Design and Implementation (PLDI 08), Tucson, AZ (June 2008)}}. Citeseer.
\newblock


\bibitem[Bondhugula et~al\mbox{.}(2008b)]%
        {bondhugula2008practical}
\bibfield{author}{\bibinfo{person}{Uday Bondhugula}, \bibinfo{person}{Albert Hartono}, \bibinfo{person}{Jagannathan Ramanujam}, {and} \bibinfo{person}{Ponnuswamy Sadayappan}.} \bibinfo{year}{2008}\natexlab{b}.
\newblock \showarticletitle{A practical automatic polyhedral parallelizer and locality optimizer}. In \bibinfo{booktitle}{\emph{Proceedings of the 29th ACM SIGPLAN Conference on Programming Language Design and Implementation}}. \bibinfo{pages}{101--113}.
\newblock


\bibitem[Bondhugula et~al\mbox{.}(2008c)]%
        {bondhugula_practical_2008}
\bibfield{author}{\bibinfo{person}{Uday Bondhugula}, \bibinfo{person}{Albert Hartono}, \bibinfo{person}{J. Ramanujam}, {and} \bibinfo{person}{P. Sadayappan}.} \bibinfo{year}{2008}\natexlab{c}.
\newblock \showarticletitle{A practical automatic polyhedral parallelizer and locality optimizer}. In \bibinfo{booktitle}{\emph{PLDI}}. \bibinfo{pages}{101--113}.
\newblock


\bibitem[Brauckmann et~al\mbox{.}(2021)]%
        {brauckmann2021reinforcement}
\bibfield{author}{\bibinfo{person}{Alexander Brauckmann}, \bibinfo{person}{Andr{\'e}s Goens}, {and} \bibinfo{person}{Jeronimo Castrillon}.} \bibinfo{year}{2021}\natexlab{}.
\newblock \showarticletitle{A reinforcement learning environment for polyhedral optimizations}.
\newblock \bibinfo{journal}{\emph{arXiv preprint arXiv:2104.13732}} (\bibinfo{year}{2021}).
\newblock


\bibitem[Brody et~al\mbox{.}(2022)]%
        {brody2022attentive}
\bibfield{author}{\bibinfo{person}{Shaked Brody}, \bibinfo{person}{Uri Alon}, {and} \bibinfo{person}{Eran Yahav}.} \bibinfo{year}{2022}\natexlab{}.
\newblock \bibinfo{title}{How Attentive are Graph Attention Networks?}
\newblock
\newblock
\showeprint[arxiv]{2105.14491}~[cs.LG]


\bibitem[Chen et~al\mbox{.}(2018)]%
        {chen2018tvm}
\bibfield{author}{\bibinfo{person}{Tianqi Chen}, \bibinfo{person}{Thierry Moreau}, \bibinfo{person}{Ziheng Jiang}, \bibinfo{person}{Lianmin Zheng}, \bibinfo{person}{Eddie Yan}, \bibinfo{person}{Meghan Cowan}, \bibinfo{person}{Haichen Shen}, \bibinfo{person}{Leyuan Wang}, \bibinfo{person}{Yuwei Hu}, \bibinfo{person}{Luis Ceze}, {et~al\mbox{.}}} \bibinfo{year}{2018}\natexlab{}.
\newblock \showarticletitle{TVM: An automated end-to-end optimizing compiler for deep learning}.
\newblock \bibinfo{journal}{\emph{arXiv preprint arXiv:1802.04799}} (\bibinfo{year}{2018}).
\newblock


\bibitem[Chen et~al\mbox{.}(2019)]%
        {chen2019learning}
\bibfield{author}{\bibinfo{person}{Tianqi Chen}, \bibinfo{person}{Lianmin Zheng}, \bibinfo{person}{Eddie Yan}, \bibinfo{person}{Ziheng Jiang}, \bibinfo{person}{Thierry Moreau}, \bibinfo{person}{Luis Ceze}, \bibinfo{person}{Carlos Guestrin}, {and} \bibinfo{person}{Arvind Krishnamurthy}.} \bibinfo{year}{2019}\natexlab{}.
\newblock \bibinfo{title}{Learning to Optimize Tensor Programs}.
\newblock
\newblock
\showeprint[arxiv]{1805.08166}


\bibitem[Cummins et~al\mbox{.}(2022)]%
        {cummins2022compilergym}
\bibfield{author}{\bibinfo{person}{Chris Cummins}, \bibinfo{person}{Bram Wasti}, \bibinfo{person}{Jiadong Guo}, \bibinfo{person}{Brandon Cui}, \bibinfo{person}{Jason Ansel}, \bibinfo{person}{Sahir Gomez}, \bibinfo{person}{Somya Jain}, \bibinfo{person}{Jia Liu}, \bibinfo{person}{Olivier Teytaud}, \bibinfo{person}{Benoit Steiner}, {et~al\mbox{.}}} \bibinfo{year}{2022}\natexlab{}.
\newblock \showarticletitle{Compilergym: Robust, performant compiler optimization environments for ai research}. In \bibinfo{booktitle}{\emph{2022 IEEE/ACM International Symposium on Code Generation and Optimization (CGO)}}. IEEE, \bibinfo{pages}{92--105}.
\newblock


\bibitem[Darte and Huard(2005)]%
        {Darte_contraction_2005}
\bibfield{author}{\bibinfo{person}{Alain Darte} {and} \bibinfo{person}{Guillaume Huard}.} \bibinfo{year}{2005}\natexlab{}.
\newblock \showarticletitle{New Complexity Results on Array Contraction and Related Problems}.
\newblock \bibinfo{journal}{\emph{J. VLSI Signal Process. Syst.}} \bibinfo{volume}{40}, \bibinfo{number}{1} (\bibinfo{date}{May} \bibinfo{year}{2005}), \bibinfo{pages}{35--55}.
\newblock
\showISSN{0922-5773}
\urldef\tempurl%
\url{https://doi.org/10.1007/s11265-005-4937-3}
\showDOI{\tempurl}


\bibitem[Feautrier(1988)]%
        {feautrier_array_1988}
\bibfield{author}{\bibinfo{person}{P. Feautrier}.} \bibinfo{year}{1988}\natexlab{}.
\newblock \showarticletitle{Array expansion}. In \bibinfo{booktitle}{\emph{Proceedings of the 2nd international conference on Supercomputing}}. \bibinfo{publisher}{{ACM}}, \bibinfo{address}{St. Malo, France}, \bibinfo{pages}{429--441}.
\newblock
\showISBNx{0-89791-272-1}
\urldef\tempurl%
\url{https://doi.org/10.1145/55364.55406}
\showDOI{\tempurl}


\bibitem[Feautrier and Lengauer(2011)]%
        {Feautrier2011}
\bibfield{author}{\bibinfo{person}{Paul Feautrier} {and} \bibinfo{person}{Christian Lengauer}.} \bibinfo{year}{2011}\natexlab{}.
\newblock \bibinfo{booktitle}{\emph{Polyhedron Model}}.
\newblock \bibinfo{publisher}{Springer US}, \bibinfo{address}{Boston, MA}, \bibinfo{pages}{1581--1592}.
\newblock
\showISBNx{978-0-387-09766-4}
\urldef\tempurl%
\url{https://doi.org/10.1007/978-0-387-09766-4_502}
\showDOI{\tempurl}


\bibitem[Gilmer et~al\mbox{.}(2017)]%
        {gilmer2017neural}
\bibfield{author}{\bibinfo{person}{Justin Gilmer}, \bibinfo{person}{Samuel~S Schoenholz}, \bibinfo{person}{Patrick~F Riley}, \bibinfo{person}{Oriol Vinyals}, {and} \bibinfo{person}{George~E Dahl}.} \bibinfo{year}{2017}\natexlab{}.
\newblock \showarticletitle{Neural message passing for quantum chemistry}. In \bibinfo{booktitle}{\emph{International conference on machine learning}}. PMLR, \bibinfo{pages}{1263--1272}.
\newblock


\bibitem[Grosser et~al\mbox{.}(2014)]%
        {tobias_hexagonal_cgo13}
\bibfield{author}{\bibinfo{person}{Tobias Grosser}, \bibinfo{person}{Albert Cohen}, \bibinfo{person}{Justin Holewinski}, \bibinfo{person}{P. Sadayappan}, {and} \bibinfo{person}{Sven Verdoolaege}.} \bibinfo{year}{2014}\natexlab{}.
\newblock \showarticletitle{Hybrid Hexagonal/Classical Tiling for GPUs}. In \bibinfo{booktitle}{\emph{Proceedings of Annual IEEE/ACM International Symposium on Code Generation and Optimization}} (Orlando, FL, USA) \emph{(\bibinfo{series}{CGO '14})}. \bibinfo{publisher}{ACM}, \bibinfo{address}{New York, NY, USA}, Article \bibinfo{articleno}{66}, \bibinfo{numpages}{10}~pages.
\newblock


\bibitem[Grosser et~al\mbox{.}(2012)]%
        {polly}
\bibfield{author}{\bibinfo{person}{Tobias Grosser}, \bibinfo{person}{Armin Groslinger}, {and} \bibinfo{person}{Christian Lengauer}.} \bibinfo{year}{2012}\natexlab{}.
\newblock \showarticletitle{Polly - Performing Polyhedral Optimizations on a Low-Level Intermediate Representation.}
\newblock \bibinfo{journal}{\emph{Parallel Processing Letters}} \bibinfo{volume}{22}, \bibinfo{number}{4} (\bibinfo{year}{2012}).
\newblock
\urldef\tempurl%
\url{http://dblp.uni-trier.de/db/journals/ppl/ppl22.html#GrosserGL12}
\showURL{%
\tempurl}


\bibitem[Haj-Ali et~al\mbox{.}(2020)]%
        {haj2020protuner}
\bibfield{author}{\bibinfo{person}{Ameer Haj-Ali}, \bibinfo{person}{Hasan Genc}, \bibinfo{person}{Qijing Huang}, \bibinfo{person}{William Moses}, \bibinfo{person}{John Wawrzynek}, \bibinfo{person}{Krste Asanovi{\'c}}, {and} \bibinfo{person}{Ion Stoica}.} \bibinfo{year}{2020}\natexlab{}.
\newblock \showarticletitle{Protuner: tuning programs with monte carlo tree search}.
\newblock \bibinfo{journal}{\emph{arXiv preprint arXiv:2005.13685}} (\bibinfo{year}{2020}).
\newblock


\bibitem[Hakimi et~al\mbox{.}(2023)]%
        {hakimi2023hybrid}
\bibfield{author}{\bibinfo{person}{Yacine Hakimi}, \bibinfo{person}{Riyadh Baghdadi}, {and} \bibinfo{person}{Yacine Challal}.} \bibinfo{year}{2023}\natexlab{}.
\newblock \showarticletitle{A hybrid machine learning model for code optimization}.
\newblock \bibinfo{journal}{\emph{International Journal of Parallel Programming}} \bibinfo{volume}{51}, \bibinfo{number}{6} (\bibinfo{year}{2023}), \bibinfo{pages}{309--331}.
\newblock


\bibitem[Hamilton et~al\mbox{.}(2017)]%
        {hamilton2017inductive}
\bibfield{author}{\bibinfo{person}{Will Hamilton}, \bibinfo{person}{Zhitao Ying}, {and} \bibinfo{person}{Jure Leskovec}.} \bibinfo{year}{2017}\natexlab{}.
\newblock \showarticletitle{Inductive representation learning on large graphs}.
\newblock \bibinfo{journal}{\emph{Advances in neural information processing systems}}  \bibinfo{volume}{30} (\bibinfo{year}{2017}).
\newblock


\bibitem[He et~al\mbox{.}(2023)]%
        {he2023xrlflowgraphreinforcementlearning}
\bibfield{author}{\bibinfo{person}{Guoliang He}, \bibinfo{person}{Sean Parker}, {and} \bibinfo{person}{Eiko Yoneki}.} \bibinfo{year}{2023}\natexlab{}.
\newblock \bibinfo{title}{X-RLflow: Graph Reinforcement Learning for Neural Network Subgraphs Transformation}.
\newblock
\newblock
\showeprint[arxiv]{2304.14698}~[cs.LG]
\urldef\tempurl%
\url{https://arxiv.org/abs/2304.14698}
\showURL{%
\tempurl}


\bibitem[Huang et~al\mbox{.}(2020)]%
        {huang2020autophase}
\bibfield{author}{\bibinfo{person}{Qijing Huang}, \bibinfo{person}{Ameer Haj-Ali}, \bibinfo{person}{William Moses}, \bibinfo{person}{John Xiang}, \bibinfo{person}{Ion Stoica}, \bibinfo{person}{Krste Asanovic}, {and} \bibinfo{person}{John Wawrzynek}.} \bibinfo{year}{2020}\natexlab{}.
\newblock \showarticletitle{Autophase: Juggling hls phase orderings in random forests with deep reinforcement learning}.
\newblock \bibinfo{journal}{\emph{arXiv preprint arXiv:2003.00671}} (\bibinfo{year}{2020}).
\newblock


\bibitem[Huanting et~al\mbox{.}(2022)]%
        {huanting2022ss}
\bibfield{author}{\bibinfo{person}{Wang Huanting}, \bibinfo{person}{Tang Zhanyong}, \bibinfo{person}{Zhang Cheng}, \bibinfo{person}{Zhao Jiaqi}, \bibinfo{person}{Cummins Chris}, \bibinfo{person}{Leather Hugh}, {and} \bibinfo{person}{Wang Zheng}.} \bibinfo{year}{2022}\natexlab{}.
\newblock \showarticletitle{Automating Reinforcement Learning Architecture Design for Code Optimization}. In \bibinfo{booktitle}{\emph{Proceedings of the 31st ACM SIGPLAN International Conference on Compiler Construction}} (Seoul, South Korea) \emph{(\bibinfo{series}{CC 2022})}. \bibinfo{publisher}{Association for Computing Machinery}, \bibinfo{address}{New York, NY, USA}, \bibinfo{pages}{129–143}.
\newblock
\showISBNx{9781450391832}
\urldef\tempurl%
\url{https://doi.org/10.1145/3497776.3517769}
\showDOI{\tempurl}


\bibitem[Irigoin and Triolet(1988)]%
        {Iri88}
\bibfield{author}{\bibinfo{person}{F. Irigoin} {and} \bibinfo{person}{R. Triolet}.} \bibinfo{year}{1988}\natexlab{}.
\newblock \showarticletitle{Supernode Partitioning}. In \bibinfo{booktitle}{\emph{(POPL'88)}}. \bibinfo{address}{San Diego, CA}, \bibinfo{pages}{319--328}.
\newblock


\bibitem[Kipf and Welling(2016)]%
        {kipf2016semi}
\bibfield{author}{\bibinfo{person}{Thomas~N Kipf} {and} \bibinfo{person}{Max Welling}.} \bibinfo{year}{2016}\natexlab{}.
\newblock \showarticletitle{Semi-supervised classification with graph convolutional networks}.
\newblock \bibinfo{journal}{\emph{arXiv preprint arXiv:1609.02907}} (\bibinfo{year}{2016}).
\newblock


\bibitem[Klambauer et~al\mbox{.}(2017)]%
        {klambauer2017self}
\bibfield{author}{\bibinfo{person}{G{\"u}nter Klambauer}, \bibinfo{person}{Thomas Unterthiner}, \bibinfo{person}{Andreas Mayr}, {and} \bibinfo{person}{Sepp Hochreiter}.} \bibinfo{year}{2017}\natexlab{}.
\newblock \showarticletitle{Self-normalizing neural networks}.
\newblock \bibinfo{journal}{\emph{Advances in neural information processing systems}}  \bibinfo{volume}{30} (\bibinfo{year}{2017}).
\newblock


\bibitem[Lefebvre and Feautrier(1998)]%
        {lefebvre_automatic_1998}
\bibfield{author}{\bibinfo{person}{Vincent Lefebvre} {and} \bibinfo{person}{Paul Feautrier}.} \bibinfo{year}{1998}\natexlab{}.
\newblock \showarticletitle{Automatic storage management for parallel programs}.
\newblock \bibinfo{journal}{\emph{Parallel Comput.}}  \bibinfo{volume}{24} (\bibinfo{year}{1998}), \bibinfo{pages}{649--671}.
\newblock
\showISSN{01678191}
\urldef\tempurl%
\url{https://doi.org/10.1016/S0167-8191(98)00029-5}
\showDOI{\tempurl}


\bibitem[Liu and Baghdadi(2025)]%
        {liu2025data}
\bibfield{author}{\bibinfo{person}{Chunting Liu} {and} \bibinfo{person}{Riyadh Baghdadi}.} \bibinfo{year}{2025}\natexlab{}.
\newblock \showarticletitle{Data-Efficient Performance Modeling via Pre-training}. In \bibinfo{booktitle}{\emph{Proceedings of the 34th ACM SIGPLAN International Conference on Compiler Construction}}. \bibinfo{pages}{48--59}.
\newblock


\bibitem[Liu et~al\mbox{.}(2019)]%
        {liu2019optimizing}
\bibfield{author}{\bibinfo{person}{Yizhi Liu}, \bibinfo{person}{Yao Wang}, \bibinfo{person}{Ruofei Yu}, \bibinfo{person}{Mu Li}, \bibinfo{person}{Vin Sharma}, {and} \bibinfo{person}{Yida Wang}.} \bibinfo{year}{2019}\natexlab{}.
\newblock \showarticletitle{Optimizing $\{$CNN$\}$ model inference on $\{$CPUs$\}$}. In \bibinfo{booktitle}{\emph{2019 USENIX Annual Technical Conference (USENIX ATC 19)}}. \bibinfo{pages}{1025--1040}.
\newblock


\bibitem[Looks et~al\mbox{.}(2017)]%
        {looks2017deep}
\bibfield{author}{\bibinfo{person}{Moshe Looks}, \bibinfo{person}{Marcello Herreshoff}, \bibinfo{person}{DeLesley Hutchins}, {and} \bibinfo{person}{Peter Norvig}.} \bibinfo{year}{2017}\natexlab{}.
\newblock \bibinfo{title}{Deep Learning with Dynamic Computation Graphs}.
\newblock
\newblock
\showeprint[arxiv]{1702.02181}


\bibitem[Merouani et~al\mbox{.}(2024)]%
        {merouani2024looper}
\bibfield{author}{\bibinfo{person}{Massinissa Merouani}, \bibinfo{person}{Khaled~Afif Boudaoud}, \bibinfo{person}{Iheb~Nassim Aouadj}, \bibinfo{person}{Nassim Tchoulak}, \bibinfo{person}{Islem~Kara Bernou}, \bibinfo{person}{Hamza Benyamina}, \bibinfo{person}{Fatima Benbouzid-Si Tayeb}, \bibinfo{person}{Karima Benatchba}, \bibinfo{person}{Hugh Leather}, {and} \bibinfo{person}{Riyadh Baghdadi}.} \bibinfo{year}{2024}\natexlab{}.
\newblock \showarticletitle{LOOPer: A Learned Automatic Code Optimizer For Polyhedral Compilers}.
\newblock \bibinfo{journal}{\emph{arXiv preprint arXiv:2403.11522}} (\bibinfo{year}{2024}).
\newblock


\bibitem[Merouani et~al\mbox{.}(2020)]%
        {merouani2020deep}
\bibfield{author}{\bibinfo{person}{Massinissa Merouani}, \bibinfo{person}{Mohamed-Hicham Leghettas}, \bibinfo{person}{Riyadh Baghdadi}, \bibinfo{person}{Taha Arbaoui}, {and} \bibinfo{person}{Karima Benatchba}.} \bibinfo{year}{2020}\natexlab{}.
\newblock \emph{\bibinfo{title}{A deep learning based cost model for automatic code optimization in tiramisu}}.
\newblock \bibinfo{thesistype}{Ph.\,D. Dissertation}. \bibinfo{school}{PhD thesis, 10 2020}.
\newblock


\bibitem[Mezdour et~al\mbox{.}(2023)]%
        {mezdour2023deep}
\bibfield{author}{\bibinfo{person}{Lina Mezdour}, \bibinfo{person}{Khadidja Kadem}, \bibinfo{person}{Massinissa Merouani}, \bibinfo{person}{Amina~Selma Haichour}, \bibinfo{person}{Saman Amarasinghe}, {and} \bibinfo{person}{Riyadh Baghdadi}.} \bibinfo{year}{2023}\natexlab{}.
\newblock \showarticletitle{A deep learning model for loop interchange}. In \bibinfo{booktitle}{\emph{Proceedings of the 32nd ACM SIGPLAN International Conference on Compiler Construction}}. \bibinfo{pages}{50--60}.
\newblock


\bibitem[Mullapudi et~al\mbox{.}(2016)]%
        {mallupudi2016halide}
\bibfield{author}{\bibinfo{person}{Ravi~Teja Mullapudi}, \bibinfo{person}{Andrew Adams}, \bibinfo{person}{Dillon Sharlet}, \bibinfo{person}{Jonathan Ragan-Kelley}, {and} \bibinfo{person}{Kayvon Fatahalian}.} \bibinfo{year}{2016}\natexlab{}.
\newblock \showarticletitle{Automatically Scheduling Halide Image Processing Pipelines}.
\newblock \bibinfo{journal}{\emph{ACM Trans. Graph.}} \bibinfo{volume}{35}, \bibinfo{number}{4}, Article \bibinfo{articleno}{83} (\bibinfo{date}{jul} \bibinfo{year}{2016}), \bibinfo{numpages}{11}~pages.
\newblock
\showISSN{0730-0301}
\urldef\tempurl%
\url{https://doi.org/10.1145/2897824.2925952}
\showDOI{\tempurl}


\bibitem[Paliwal et~al\mbox{.}(2020)]%
        {paliwal2020reinforced}
\bibfield{author}{\bibinfo{person}{Aditya Paliwal}, \bibinfo{person}{Felix Gimeno}, \bibinfo{person}{Vinod Nair}, \bibinfo{person}{Yujia Li}, \bibinfo{person}{Miles Lubin}, \bibinfo{person}{Pushmeet Kohli}, {and} \bibinfo{person}{Oriol Vinyals}.} \bibinfo{year}{2020}\natexlab{}.
\newblock \bibinfo{title}{Reinforced Genetic Algorithm Learning for Optimizing Computation Graphs}.
\newblock
\newblock
\showeprint[arxiv]{1905.02494}


\bibitem[Pecenin et~al\mbox{.}(2019)]%
        {pecenin2019optimization}
\bibfield{author}{\bibinfo{person}{Marcelo Pecenin}, \bibinfo{person}{Andr{\'e}~Murbach Maidl}, {and} \bibinfo{person}{Daniel Weingaertner}.} \bibinfo{year}{2019}\natexlab{}.
\newblock \showarticletitle{Optimization of halide image processing schedules with reinforcement learning}. In \bibinfo{booktitle}{\emph{Anais do XX Simp{\'o}sio em Sistemas Computacionais de Alto Desempenho}}. SBC, \bibinfo{pages}{37--48}.
\newblock


\bibitem[Pouchet et~al\mbox{.}(2012)]%
        {pouchet2012polybench}
\bibfield{author}{\bibinfo{person}{Louis-No{\"e}l Pouchet} {et~al\mbox{.}}} \bibinfo{year}{2012}\natexlab{}.
\newblock \showarticletitle{Polybench: The polyhedral benchmark suite}.
\newblock \bibinfo{journal}{\emph{URL: http://www. cs. ucla. edu/pouchet/software/polybench}}  \bibinfo{volume}{437} (\bibinfo{year}{2012}), \bibinfo{pages}{1--1}.
\newblock


\bibitem[Pouchet et~al\mbox{.}(2011)]%
        {pouchet.11.popl}
\bibfield{author}{\bibinfo{person}{Louis-No{\"e}l Pouchet}, \bibinfo{person}{Uday Bondhugula}, \bibinfo{person}{C{\'e}dric Bastoul}, \bibinfo{person}{Albert Cohen}, \bibinfo{person}{J. Ramanujam}, \bibinfo{person}{P. Sadayappan}, {and} \bibinfo{person}{Nicolas Vasilache}.} \bibinfo{year}{2011}\natexlab{}.
\newblock \showarticletitle{Loop Transformations: Convexity, Pruning and Optimization}. In \bibinfo{booktitle}{\emph{38th ACM SIGACT-SIGPLAN Symposium on Principles of Programming Languages (POPL'11)}}. \bibinfo{publisher}{ACM Press}, \bibinfo{address}{Austin, TX}, \bibinfo{pages}{549--562}.
\newblock


\bibitem[Quiller\'e and Rajopadhye(2000)]%
        {Qui00}
\bibfield{author}{\bibinfo{person}{F. Quiller\'e} {and} \bibinfo{person}{S. Rajopadhye}.} \bibinfo{year}{2000}\natexlab{}.
\newblock \showarticletitle{Optimizing Memory Usage in the Polyhedral Model}.
\newblock \bibinfo{journal}{\emph{ACM Trans. on Programming Languages and Systems}} \bibinfo{volume}{22}, \bibinfo{number}{5} (\bibinfo{date}{Sept.} \bibinfo{year}{2000}), \bibinfo{pages}{773--815}.
\newblock


\bibitem[Ragan-Kelley et~al\mbox{.}(2013)]%
        {ragan2013halide}
\bibfield{author}{\bibinfo{person}{Jonathan Ragan-Kelley}, \bibinfo{person}{Connelly Barnes}, \bibinfo{person}{Andrew Adams}, \bibinfo{person}{Sylvain Paris}, \bibinfo{person}{Fr\'{e}do Durand}, {and} \bibinfo{person}{Saman Amarasinghe}.} \bibinfo{year}{2013}\natexlab{}.
\newblock \showarticletitle{Halide: A Language and Compiler for Optimizing Parallelism, Locality, and Recomputation in Image Processing Pipelines}.
\newblock \bibinfo{journal}{\emph{SIGPLAN Not.}} \bibinfo{volume}{48}, \bibinfo{number}{6} (\bibinfo{date}{jun} \bibinfo{year}{2013}), \bibinfo{pages}{519–530}.
\newblock
\showISSN{0362-1340}
\urldef\tempurl%
\url{https://doi.org/10.1145/2499370.2462176}
\showDOI{\tempurl}


\bibitem[Schulman et~al\mbox{.}(2017)]%
        {schulman2017proximal}
\bibfield{author}{\bibinfo{person}{John Schulman}, \bibinfo{person}{Filip Wolski}, \bibinfo{person}{Prafulla Dhariwal}, \bibinfo{person}{Alec Radford}, {and} \bibinfo{person}{Oleg Klimov}.} \bibinfo{year}{2017}\natexlab{}.
\newblock \showarticletitle{Proximal policy optimization algorithms}.
\newblock \bibinfo{journal}{\emph{arXiv preprint arXiv:1707.06347}} (\bibinfo{year}{2017}).
\newblock


\bibitem[Sutton and Barto(2018)]%
        {sutton2018reinforcement}
\bibfield{author}{\bibinfo{person}{R.S. Sutton} {and} \bibinfo{person}{A.G. Barto}.} \bibinfo{year}{2018}\natexlab{}.
\newblock \bibinfo{booktitle}{\emph{Reinforcement Learning, second edition: An Introduction}}.
\newblock \bibinfo{publisher}{MIT Press}.
\newblock
\showISBNx{9780262039246}
\showLCCN{2018023826}
\urldef\tempurl%
\url{https://books.google.dz/books?id=sWV0DwAAQBAJ}
\showURL{%
\tempurl}


\bibitem[Thies et~al\mbox{.}(2001)]%
        {thies_unified_2001}
\bibfield{author}{\bibinfo{person}{William Thies}, \bibinfo{person}{Fr\'{e}d\'{e}ric Vivien}, \bibinfo{person}{Jeffrey Sheldon}, {and} \bibinfo{person}{Saman Amarasinghe}.} \bibinfo{year}{2001}\natexlab{}.
\newblock \showarticletitle{A unified framework for schedule and storage optimization}. In \bibinfo{booktitle}{\emph{Proc.\ of the 2001 {PLDI} Conf.}}
\newblock


\bibitem[Trifunovic et~al\mbox{.}(2010)]%
        {trifunovic_graphite_2010}
\bibfield{author}{\bibinfo{person}{Konrad Trifunovic}, \bibinfo{person}{Albert Cohen}, \bibinfo{person}{David Edelsohn}, \bibinfo{person}{Feng Li}, \bibinfo{person}{Tobias Grosser}, \bibinfo{person}{Harsha Jagasia}, \bibinfo{person}{Razya Ladelsky}, \bibinfo{person}{Sebastian Pop}, \bibinfo{person}{Jan Sjodin}, {and} \bibinfo{person}{Ramakrishna Upadrasta}.} \bibinfo{year}{2010}\natexlab{}.
\newblock \bibinfo{title}{{GRAPHITE} Two Years After: First Lessons Learned From {Real-World} Polyhedral Compilation}.
\newblock
\newblock


\bibitem[Vasilache et~al\mbox{.}(2006)]%
        {violateddep}
\bibfield{author}{\bibinfo{person}{Nicolas Vasilache}, \bibinfo{person}{Cedric Bastoul}, \bibinfo{person}{Albert Cohen}, {and} \bibinfo{person}{Sylvain Girbal}.} \bibinfo{year}{2006}\natexlab{}.
\newblock \showarticletitle{Violated Dependence Analysis}. In \bibinfo{booktitle}{\emph{Proceedings of the 20th Annual International Conference on Supercomputing}} (Cairns, Queensland, Australia) \emph{(\bibinfo{series}{ICS '06})}. \bibinfo{publisher}{Association for Computing Machinery}, \bibinfo{address}{New York, NY, USA}, \bibinfo{pages}{335–344}.
\newblock
\showISBNx{1595932828}
\urldef\tempurl%
\url{https://doi.org/10.1145/1183401.1183448}
\showDOI{\tempurl}


\bibitem[Vasilache et~al\mbox{.}(2018a)]%
        {Vasilache2018TensorCF}
\bibfield{author}{\bibinfo{person}{Nicolas Vasilache}, \bibinfo{person}{Oleksandr Zinenko}, \bibinfo{person}{Theodoros Theodoridis}, \bibinfo{person}{Priya Goyal}, \bibinfo{person}{Zach DeVito}, \bibinfo{person}{William~S. Moses}, \bibinfo{person}{Sven Verdoolaege}, \bibinfo{person}{Andrew Adams}, {and} \bibinfo{person}{Albert Cohen}.} \bibinfo{year}{2018}\natexlab{a}.
\newblock \showarticletitle{Tensor Comprehensions: Framework-Agnostic High-Performance Machine Learning Abstractions}.
\newblock \bibinfo{journal}{\emph{CoRR}}  \bibinfo{volume}{abs/1802.04730} (\bibinfo{year}{2018}).
\newblock


\bibitem[Vasilache et~al\mbox{.}(2018b)]%
        {vasilache2018tensor}
\bibfield{author}{\bibinfo{person}{Nicolas Vasilache}, \bibinfo{person}{Oleksandr Zinenko}, \bibinfo{person}{Theodoros Theodoridis}, \bibinfo{person}{Priya Goyal}, \bibinfo{person}{Zachary DeVito}, \bibinfo{person}{William~S Moses}, \bibinfo{person}{Sven Verdoolaege}, \bibinfo{person}{Andrew Adams}, {and} \bibinfo{person}{Albert Cohen}.} \bibinfo{year}{2018}\natexlab{b}.
\newblock \showarticletitle{Tensor comprehensions: Framework-agnostic high-performance machine learning abstractions}.
\newblock \bibinfo{journal}{\emph{arXiv preprint arXiv:1802.04730}} (\bibinfo{year}{2018}).
\newblock


\bibitem[Verdoolaege et~al\mbox{.}(2013)]%
        {sven2013}
\bibfield{author}{\bibinfo{person}{Sven Verdoolaege}, \bibinfo{person}{Juan Carlos~Juega}, \bibinfo{person}{Albert Cohen}, \bibinfo{person}{Jos\'{e} Ignacio~G\'{o}mez}, \bibinfo{person}{Christian Tenllado}, {and} \bibinfo{person}{Francky Catthoor}.} \bibinfo{year}{2013}\natexlab{}.
\newblock \showarticletitle{Polyhedral parallel code generation for CUDA}.
\newblock \bibinfo{journal}{\emph{ACM Trans. Archit. Code Optim.}} \bibinfo{volume}{9}, \bibinfo{number}{4}, Article \bibinfo{articleno}{54} (\bibinfo{date}{jan} \bibinfo{year}{2013}), \bibinfo{numpages}{23}~pages.
\newblock
\showISSN{1544-3566}
\urldef\tempurl%
\url{https://doi.org/10.1145/2400682.2400713}
\showDOI{\tempurl}


\bibitem[Wolf and Lam(1991)]%
        {wolf1991loop}
\bibfield{author}{\bibinfo{person}{Michael~E Wolf} {and} \bibinfo{person}{Monica~S Lam}.} \bibinfo{year}{1991}\natexlab{}.
\newblock \showarticletitle{A loop transformation theory and an algorithm to maximize parallelism}.
\newblock \bibinfo{journal}{\emph{IEEE transactions on parallel and distributed systems}} \bibinfo{volume}{2}, \bibinfo{number}{4} (\bibinfo{year}{1991}), \bibinfo{pages}{452--471}.
\newblock


\bibitem[Zheng et~al\mbox{.}(2020)]%
        {zheng2020ansor}
\bibfield{author}{\bibinfo{person}{Lianmin Zheng}, \bibinfo{person}{Chengfan Jia}, \bibinfo{person}{Minmin Sun}, \bibinfo{person}{Zhao Wu}, \bibinfo{person}{Cody~Hao Yu}, \bibinfo{person}{Ameer Haj-Ali}, \bibinfo{person}{Yida Wang}, \bibinfo{person}{Jun Yang}, \bibinfo{person}{Danyang Zhuo}, \bibinfo{person}{Koushik Sen}, {et~al\mbox{.}}} \bibinfo{year}{2020}\natexlab{}.
\newblock \showarticletitle{Ansor: Generating $\{$High-Performance$\}$ tensor programs for deep learning}. In \bibinfo{booktitle}{\emph{14th USENIX symposium on operating systems design and implementation (OSDI 20)}}. \bibinfo{pages}{863--879}.
\newblock


\bibitem[Zheng et~al\mbox{.}(2023)]%
        {zheng2023ansor}
\bibfield{author}{\bibinfo{person}{Lianmin Zheng}, \bibinfo{person}{Chengfan Jia}, \bibinfo{person}{Minmin Sun}, \bibinfo{person}{Zhao Wu}, \bibinfo{person}{Cody~Hao Yu}, \bibinfo{person}{Ameer Haj-Ali}, \bibinfo{person}{Yida Wang}, \bibinfo{person}{Jun Yang}, \bibinfo{person}{Danyang Zhuo}, \bibinfo{person}{Koushik Sen}, \bibinfo{person}{Joseph~E. Gonzalez}, {and} \bibinfo{person}{Ion Stoica}.} \bibinfo{year}{2023}\natexlab{}.
\newblock \bibinfo{title}{Ansor: Generating High-Performance Tensor Programs for Deep Learning}.
\newblock
\newblock
\showeprint[arxiv]{2006.06762}


\bibitem[Zheng et~al\mbox{.}(2021)]%
        {zheng2021fusionstitching}
\bibfield{author}{\bibinfo{person}{Zhen Zheng}, \bibinfo{person}{Pengzhan Zhao}, \bibinfo{person}{Guoping Long}, \bibinfo{person}{Feiwen Zhu}, \bibinfo{person}{Kai Zhu}, \bibinfo{person}{Wenyi Zhao}, \bibinfo{person}{Lansong Diao}, \bibinfo{person}{Jun Yang}, {and} \bibinfo{person}{Wei Lin}.} \bibinfo{year}{2021}\natexlab{}.
\newblock \bibinfo{title}{FusionStitching: Boosting Memory Intensive Computations for Deep Learning Workloads}.
\newblock
\newblock
\showeprint[arxiv]{2009.10924}


\end{thebibliography}
\end{document}